\RequirePackage[2020-02-02]{latexrelease}
\documentclass[eqsecnum,noshowpacs,showkeys,nofootinbib,aps]{revtex4}

\usepackage{amsmath}
\usepackage{epsfig}
\usepackage{xcolor}
\usepackage{ulem} 

\renewcommand{\theequation}{\arabic{equation}}
\def\be{\begin{equation}}
\def\ee{\end{equation}}
\def\bea{\begin{eqnarray}}
\def\eea{\end{eqnarray}}

\def\na{\nabla}

\begin{document}

\title{GEMS embeddings of Hayward regular black holes in massless and massive
gravities}
\author{Soon-Tae Hong}
\email{galaxy.mass@gmail.com}

\affiliation{Center for Quantum
Spacetime, Sogang University, Seoul 04107, Korea}

\affiliation{Department of Physics, Sogang University, Seoul
04107, Korea}

\author{Yong-Wan Kim}
\email{ywkim65@gmail.com}

\affiliation{Research Institute of Physics and Chemistry, Jeonbuk
National University, Jeonju 54896, Korea}

\affiliation{Department of Physics, Jeonbuk National University,
Jeonju 54896, Korea}

\author{Young-Jai Park}
\affiliation{Department of Physics, Sogang University, Seoul
04107, Korea }

\date{\today}

\begin{abstract}
After finding a solution for the Hayward regular black hole (HRBH)
in massive gravity, we embed the (3+1)-dimensional HRBHs both in
massless and in massive gravities into (5+2)- and
(6+3)-dimensional Minkowski spacetimes, respectively. Here,
massive gravity denotes that a graviton acquires a mass
holographically by broken momentum conservation in the HRBH. The
original HRBH has no holographically added gravitons, which we
call massless. Making use of newly found embedding coordinates, we
obtain desired Unruh temperatures and compare them with the
Hawking and local fiducial temperatures, showing that the Unruh
effect for a uniformly accelerated observer in a higher
dimensional flat spacetime is equal to the Hawking effect for a
fiducial observer in a black hole spacetime. We also obtain freely
falling temperatures of the HRBHs in massless and massive
gravities seen by freely falling observers, which remain finite
even at the event horizons while becoming the Hawking temperatures
in asymptotic infinity.
\end{abstract}
\pacs{04.20.-q, 04.50.-h, 04.70.-s}

\keywords{Hayward regular black hole; global flat embedding; Unruh
effect; freely falling temperature}

\maketitle

\section{introduction}
\setcounter{equation}{0}
\renewcommand{\theequation}{\arabic{section}.\arabic{equation}}

Black holes are among the most mysterious and fascinating objects
in our universe, observationally as well as theoretically.
Theoretically, it dates back to a century ago when Schwarzschild
found a spherically symmetric vacuum solution of Einstein's
gravitational field equations, and in the same year, Einstein
suggested the existence of gravitational waves as a natural
outcome of his general relativity. Observationally, precisely a
century later, LIGO with Virgo collaboration
\cite{LIGOScientific:2016aoc} detected the gravitational waves,
representing the merger of two stellar mass black holes. In 2019,
the Event Horizon Telescope (EHT)
\cite{EventHorizonTelescope:2019dse} finally detected the direct
image of a black hole and its event horizon. These remarkable
discoveries provide some constraints on the modified theories of
gravity, such as the existence of a tight bound on the graviton
mass \cite{LIGOScientific:2016aoc}, the speed of the gravitational
wave \cite{Cornish:2017jml}, black hole parameters and properties
\cite{Zhang:2017jze,deRham:2019wjj,Vagnozzi:2022moj}, and so on,
in the light of observational data.

The generalization of the Schwarzschild black hole solution to the
electrically charged one \cite{Reissner:1916,Nordstrom:2018} was
done immediately after the work, and the more general and
realistic rotating black hole solutions
\cite{Kerr:1963,Newman:1965} were found in the 1960s. However, all
of these black hole solutions have a curvature singularity at $r =
0$ in which spacetime is geodesically incomplete. According to the
singularity theorem proved by Hawking and Penrose
\cite{Hawking:1973}, singularity is inevitable. Moreover, the
usual laws of physics break down at the singularity and many
physicists believe that quantum gravity effects would work near
the singularity. Nevertheless, since we do not have a complete
quantum gravity theory yet, another line of work to avoid
singularity has been pursued on regular black holes. Such works
began with Gliner \cite{Gliner:1966} and Sakharov
\cite{Sakharov:1966} who proposed a way to avoid the singularity
in terms of the matter source, which has a de Sitter core with an
equation of state $\rho=-p$ at the center of the spacetime. To
avoid the singularity problem, Bardeen \cite{Bardeen:1968} also
proposed a model of a regular black hole obeying the weak energy
condition and thus the model does not obey at least one condition
of Hawking-Penrose's singularity theorem. However, the Bardeen's
regular black hole solution is not an exact solution to Einstein's
equations. After more than three decades, Ayon-Beato and Garcia
\cite{Ayon-Beato:2000mjt} showed that the Bardeen's metric could
be interpreted as a magnetic solution to Einstein's equations
coupled to nonlinear electrodynamics. Since then, there has been a
lot of work on regular black holes, including Dymnikova
\cite{Dymnikova:1992ux,Dymnikova:1999cz},
Bronnikov\cite{Bronnikov:2000vy}, Hayward \cite{Hayward:2005gi},
Ayon-Beato and Garcia \cite{Ayon-Beato:1998hmi}, and more
\cite{Mars:1996khm,Borde:1996df,Burinskii:2002pz,Mbonye:2005im,Bronnikov:2005gm,Berej:2006cc,
Bambi:2013ufa,Balart:2014cga,Balart:2016zrd,Fan:2016hvf}. These
studies have inspired further investigations related to such black
holes, for example, regarding particle geodesics
\cite{Abbas:2014oua,Stuchlik:2014qja,Zhao:2017cwk,Chiba:2017nml,Pradhan:2018oha,
Dymnikova:2019muu,Zhang:2021uaz,Khan:2021wzm,Rayimbaev:2022hrn},
the shadows of regular black holes
\cite{Lamy:2018zvj,Stuchlik:2019uvf,Becerril:2020fek,Ling:2022vrv},
and the quasi-normal modes
\cite{Fernando:2012yw,Flachi:2012nv,Lin:2013ofa,Saleh:2018hba,Cai:2021ele}.
The thermodynamics and phase transitions for regular black holes
have also been studied widely in Refs.
\cite{Myung:2007av,Sharif:2011ja,Saadat:2013rna,Tharanath:2014naa,
Sebastian:2015psa,Gan:2016pbu,Ghosh:2018bxg,Ali:2018boy,Aros:2019quj,
Nam:2019clw,Kruglov:2019wjv,Kumar:2020uyz,Kumar:2020xvu,Merriam:2021bar,Sharif:2022oym}.

On the other hand, as is well known, Einstein's general relativity
(GR) is a theory of a massless graviton. However, quantum gravity
phenomenology \cite{AmelinoCamelia:2008qg} at extreme limits has
pushed forward to search for alternatives to GR, one of which is
to introduce a massive graviton to GR. Historically, it started
with Fierz and Pauli \cite{Fierz:1939ix} who developed a massive
theory by extending GR with a quadratic mass term. However, the
theory suffers from the Boulware-Deser ghost problem
\cite{Boulware:1973my} and the van Dam, Veltman, and Zakharov
(vDVZ) discontinuity \cite{vanDam:1970vg,Zakharov:1970cc} in the
massless graviton limit. The vDVZ discontinuity was cured by the
Vainshtein mechanism \cite{Vainshtein:1972sx}. After half a
century, the notorious Boulware-Deser ghost problem was at last
solved by de Rham, Gabadadze, and Trolley (dRGT)
\cite{deRham:2010ik,deRham:2010kj} to have a ghost free massive
gravity, which has nonlinearly interacting mass terms constructed
from the metric coupled with a symmetric reference metric tensor.
These new terms with properly tuned coefficients make it avoid the
ghosts order by order. To all orders, the complete absence of the
Boulware-Deser ghost was subsequently proven by Hassan and Rosen
by a Hamiltonian analysis of the untruncated theory
\cite{Hassan:2011hr,Hassan:2011tf} and other works
\cite{Kluson:2011qe,Kluson:2011rt,Kluson:2012gz,Comelli:2012vz,Golovnev:2011aa,Deffayet:2012nr}.
Since then, the dRGT massive gravity has led to new astronomical
and cosmological applications for modified gravity
\cite{Ghosh:2015cva,Arraut:2014uza,Panpanich:2018cxo,
Hou:2020yni,Akbarieh:2021vhv,Aslmarand:2021qwn,
Akbarieh:2022ovn,Kazempour:2022asl,Kazempour:2022let,Kazempour:2022giy}
including the black hole shadow
\cite{Panpanich:2019mll,Upadhyay:2023yhk}.In particular, Hendi et
al. \cite{Hendi:2022qgi} have obtained a fascinating result of
having the allowed regions of the massive parameters by comparing
the black hole shadow in the dRGT massive gravity with the EHT
data of M87$^*$. On the other hand, Vegh \cite{Vegh:2013sk}
further elaborated on another nonlinear massive gravity with a
special singular reference metric to apply it to gauge/gravity
duality. The modification in the reference metric in the dRGT
massive gravity keeps the diffeomorphism symmetry for coordinates
($t,r$) intact but breaks it in angular directions so that
gravitons acquire the mass because of broken momentum conservation
\cite{Davison:2013jba,Blake:2013bqa,Blake:2013owa}. As a result,
momentum dissipates as the graviton may behave like a lattice, and
it can avoid divergent conductivity. Since then, this Vegh's type
of massive gravity, called holographic massive gravity, has been
extensively exploited to investigate many interesting models in
gravity
\cite{Cai:2014znn,Adams:2014vza,Hendi:2015pda,Hu:2016hpm,Zou:2016sab,
Hendi:2017fxp,Tannukij:2017jtn,Hendi:2017ibm,Hendi:2017bys,
EslamPanah:2018evk,Hendi:2018xuy,Chabab:2019mlu,EslamPanah:2019fci}
Very recently, we have studied the tidal effects
\cite{Hong:2020bdb} and statistical entropy \cite{Hong:2021xeg} in
the Schwarzschild black hole in holographic massive gravity.

We have also studied the charged BTZ \cite{Hong:2018spz} and
Schwarzschild black holes \cite{Hong:2019zsi} in holographic
massive gravity in the global embedding Minkowski spacetime (GEMS)
scheme. According to the GEMS scheme, any low dimensional
Riemannian manifold can be locally isometrically embedded in a
higher dimensional flat one
\cite{Fronsdal:1959zza,Rosen:1965,Goenner1980}. This can make us
have a complete analytic extension of manifolds, or we can use it
for visualizing or deriving physical properties of the embedded
spacetimes, such as a unified description of Hawking
\cite{Hawking:1974sw} and Unruh effects \cite{Unruh:1976db}. In
this line of work, Deser and Levin
\cite{Deser:1997ri,Deser:1998bb,Deser:1998xb} firstly showed that
the Hawking temperature for a fiducial observer in a curved
spacetime can be considered as the Unruh one for a uniformly
accelerated observer in a higher-dimensional GEMS embedded flat
spacetime. Since then, there has been much work on the GEMS
approach to confirm these ideas in various other spacetimes
\cite{Hong:2000kn,Kim:2000ct,Hong:2003xz,Chen:2004qw,Santos:2004ws,
Banerjee:2010ma,Cai:2010bv,Hu:2011yx,Hong:2000bp,Hong:2005rn,
Hong:2003dj,Paston:2014efa,Sheykin:2019uwj} and an interesting
extension to embedding gravity
\cite{Paston:2011wp,Paston:2018orc,Paston:2020otr,Paston:2020isf}.
Later, Brynjolfsson and Thorlacius \cite{Brynjolfsson:2008uc}
introduced a local temperature measured by a freely falling
observer in the GEMS method. We have also studied various
interesting curved spacetimes \cite{Kim:2013wpa,Hong:2020dow} to
investigate local temperatures and their equivalence to Hawking
ones.

The main goal of this paper is to construct and analyze the GEMS
embeddings of spacetimes having regular black holes in massless
and massive gravity, whose embeddings are neither found nor even
tried at all, as far as we know. Moreover, a recent study on the
geodesic completion of a regular black hole \cite{Zhou:2022yio}
supports the needs of the GEMS embeddings of regular black holes.
In this respect, it would be interesting to embed a regular black
hole with massive gravitons into a higher dimensional flat
spacetime. In this paper, we will consider the Hayward regular
black hole (HRBH) in massless gravity as a representative of
regular black holes and extend it to massive gravity to embed it
in higher dimensional flat spacetimes. Here, the HRBH in massless
gravity means the original Hayward black hole, while the HRBH in
massive gravity is the one having massive gravitons obtained from
the consideration of Vegh's type of massive gravity. We note that
when the Hayward parameter vanishes, the HRBH in massless gravity
becomes the Schwarzschild black hole, and when massive gravitons
are turned off, the HRBH in massive gravity is reduced to the HRBH
in massless gravity.

The remainder of the paper is organized as follows. In Sec. II, we
will newly find solutions to the HRBH in holographic massive
gravity. We will first briefly summarize the known solution of the
HRBH in massless gravity and generalize it to one in holographic
massive gravity. Then, we will show that the Kretschmann scalar
for the HRBH in massive gravity is not regular near $r=0$. In Sec.
III, we will construct the GEMS embeddings of the HRBH both in
massless and in massive gravity. As a result, making use of
embedding coordinates, we will obtain Unruh, Hawking, and freely
falling temperatures seen by different observers. Conclusions are
drawn in Sec. IV. Lastly, since embedding coordinates of these
regular black holes are very complicated, we have separately
listed them and then shown their limits explicitly from the
massive to the massless cases in Appendix A.

\section{HRBH in massive gravity}
\setcounter{equation}{0}
\renewcommand{\theequation}{\arabic{section}.\arabic{equation}}

The (3+1)-dimensional HRBH in holographic massive gravity is
described by the action
 \be\label{Hayward-action}
 S=\frac{1}{16\pi G}\int d^4x\sqrt{-g}\left[{\cal R}
   +\frac{24l^2m^2}{(r^3+2l^2m)^2}
   +{\tilde m}^2\sum_{a=1}^{4}c_a{\cal U}_{a}(g_{\mu\nu},f_{\mu\nu})\right],
 \ee
where ${\cal R}$ is the scalar curvature of the metric
$g_{\mu\nu}$, $m$ is the black hole mass, $l$ is a length-scale
Hayward parameter present in the Hayward solution, $\tilde{m}$ is
a graviton mass\footnote{In this paper, we shall call it massless
when $\tilde{m}$ is zero.}, $c_a$ are the coupling constants,
$f_{\mu\nu}$ is a fixed symmetric tensor usually called the
reference metric, and ${\cal U}_a$ are symmetric polynomial
potentials of the eigenvalue of the matrix ${\cal
K}^\mu_\nu\equiv\sqrt{g^{\mu\alpha}f_{\alpha\nu}}$ given as
 \bea
 {\cal U}_1 &=& [{\cal K}],\nonumber\\
 {\cal U}_2 &=& [{\cal K}]^2-[{\cal K}^2],\nonumber\\
 {\cal U}_3 &=& [{\cal K}]^3-3[{\cal K}][{\cal K}^2]+2[{\cal K}^3],\nonumber\\
 {\cal U}_4 &=& [{\cal K}]^4-6[{\cal K}^2][{\cal K}]^2+8[{\cal K}^3][{\cal K}]+3[{\cal K}^2]^2-6[{\cal K}^4].
 \eea
Here, the square root in ${\cal K}$ means
$(\sqrt{A})^\mu_\alpha(\sqrt{A})^\alpha_\nu=A^\mu_\nu$ and square
brackets denote the trace $[{\cal K}]={\cal K}^\mu_\mu$. Indices
are raised and lowered with the dynamical metric $g_{\mu\nu}$,
while the reference metric $f_{\mu\nu}$ is a non-dynamical, fixed
symmetric tensor that is introduced to construct nontrivial
interaction terms in holographic massive gravity.

Variation of the action (\ref{Hayward-action}) with respect to the
metric $g_{\mu\nu}$ leads to the equations of motion given by
 \begin{eqnarray}
 \label{eom}
 {\cal R}_{\mu\nu}
 &-& \frac{1}{2}g_{\mu\nu}\left({\cal R}-\frac{24l^2m^2}{(r^3+2l^2m)^2}
     +\tilde{m}^2\sum^{4}_{a=1}c_a{\cal U}_a\right)\nonumber\\
 &+&\frac{1}{2}\tilde{m}^2\sum^{4}_{a=1}\left[ac_a{\cal U}_{a-1}{\cal K}_{\mu\nu}
 -a(a-1)c_a{\cal U}_{a-2}{\cal K}^2_{\mu\nu}
 +6(3a-8)c_a{\cal U}_{a-3}{\cal K}^3_{\mu\nu}
 -12(a-2)c_a{\cal U}_{a-4}{\cal K}^4_{\mu\nu}\right]=0.
 \end{eqnarray}
with ${\cal U}_{-a}=0$ and ${\cal U}_0=1$.

When one considers the spherically symmetric black hole solution
ansatz as
 \be\label{metric-Hayward}
 ds^2=-f(r)dt^2+f^{-1}(r)dr^2+r^2(d\theta^2+\sin^2\theta d\phi^2)
 \ee
with the following degenerate reference metric
 \be\label{fidmetric}
 f_{\mu\nu}={\rm diag}(0,0,c^2_0,c^2_0\sin^2\theta),
 \ee
one can find
 \be
 {\cal K}^\theta_\theta={\cal K}^\phi_\phi=\frac{c_0}{r}.
 \ee
Note that the choice of the reference metric in Eq.
(\ref{fidmetric}) preserves general covariance in ($t,r$) but not
in the angular directions. This gives the symmetric potentials as
 \be
 {\cal U}_1=\frac{2c_0}{r},~~~{\cal U}_2=\frac{2c^2_0}{r^2},~~~{\cal U}_3={\cal
 U}_4=0.
 \ee
Therefore, we see that there are no contributions from $c_3$ and
$c_4$ terms which would appear in (4+1) and (5+1)-dimensional
spacetimes, respectively. Then, we finally obtain the new solution
of the HRBH in massive gravity as
 \be\label{lapsewol}
 f(r)=1-\frac{2mr^2}{r^3+2l^2m}+2Rr+{\cal C}
 \ee
with graviton mass parameters $R=c_0c_1\tilde{m}^2/4$ and ${\cal
C}=c^2_0c_2\tilde{m}^2$. Note that $m$ is an integration constant
related to the mass of the black hole and $c_0$ is a positive
constant. In the limit of $R\rightarrow 0$ and ${\cal
C}\rightarrow 0$, it becomes the HRBH in massless gravity, and
when $l\rightarrow 0$, it is further reduced to the Schwarzschild
metric as expected. On the other hand, when $R\neq 0$ and ${\cal
C}\neq 0$ with $l=0$, it becomes the Schwarzschild black hole in
massive gravity \cite{Hong:2019zsi}.

It is appropriate to comment on the two terms due to massive
gravitons. Firstly, the ${\cal C}$  term in Eq. (\ref{lapsewol})
reminds us of a monopole solution introduced by Barriola and
Vilenkin \cite{Barriola:1989hx}, which comes from a topological
defect in the early Universe as a result of a gauge symmetry
breaking. On the other hand, the $R$ term is not uncommon in
gravity theories, which also arises in, such as the dRGT massive
gravity \cite{deRham:2010ik,deRham:2010kj}, conformal gravity
\cite{Riegert:1984zz}, and $f(R)$ gravity \cite{Saffari:2007zt}.
Physically, the linear term in Eq. (\ref{lapsewol}) stands for a
deviation between the solution and the HRBH spacetime in massless
gravity, as in the modified Newtonian dynamics (MOND) studies a
deviation of massive bodies in the solar system from the Newtonian
mechanics \cite{Milgrom:1983ca}. The linear term also affects the
radius of the photon sphere in a black hole, the size of the
shadow, which is smaller than the one in a Schwarzschild spacetime
\cite{Gregoris:2021plc}. In addition, Hendi et al.
\cite{Hendi:2022qgi} have shown that the black hole shadow in the
dRGT massive gravity consistent with EHT data is given for small
$R<0.072$ and negative $-0.3<{\cal C}<-0.03$ by comparing the
black hole shadow with the data of EHT collaboration. We note that
the additional ${\cal C}$ and $R$ terms in Eq. (\ref{lapsewol})
are obtained from the consideration of massive gravitons, while
the other theories have different causes.

\begin{figure*}[t!]
   \centering
   \includegraphics{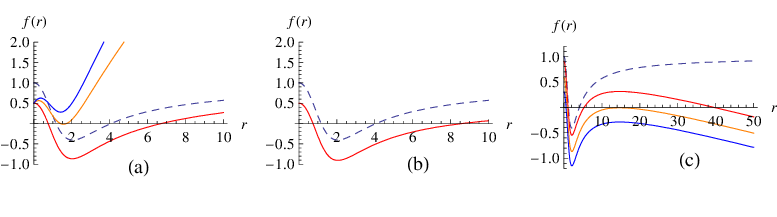}
\caption{$f(r)$-graphs for the third column case in Table
\ref{table1} of $-1<{\cal C}<0$ and $R$: (a) for $R>0$, where $(R,
{\cal C})=(0.01,-0.5),~ (0.25,-0.5),~ (0.35,-0.5)$ from top to
bottom curves, (b) $R=0$, where $(R, {\cal C})=(0,-0.5)$ and (c)
$R<0$, where $(R, {\cal
C})=(-0.01,-0.1),~(-0.01,-0.42),~(-0.01,-0.7)$ from top to bottom
curves. The dashed lines corresponding to $(R, {\cal C})=(0,0)$
are for the HRBH in massless gravity having two horizons. Here, we
have chosen $m=5\sqrt{3}/4>m_*$ and $l=1$.}
 \label{fig1}
\end{figure*}

It is well known that the HRBH solution in massless gravity ($R=0$
and ${\cal C}=0$) has two horizons for $m>m_*\equiv 3\sqrt{3}l/4$,
one for $m=m_*$, and none for $m<m_*$ \cite{Hayward:2005gi}.
However, for the case of the HRBH in massive gravity, these are
modified by $R$ and ${\cal C}$. While we will find their exact
solutions of $f(r_H)=0$ in the following subsections, we summarize
here the numbers of event horizons of the HRBH in holographic
massive gravity according to various values of $R$ and ${\cal C}$
in Tables I and II. In the Tables, max(2) or max(3) mean that
there exist maximum two or three horizons, respectively,
satisfying $f(r_H)=0$ in the given range of $R$ and ${\cal C}$.
Otherwise, a given number of event horizons are allowed in the
range. In order to describe these clearly,  as an example, in Fig.
\ref{fig1}, we have plotted $f(r)$-graphs for $-1<{\cal C}<0$ with
$R$ for the case of $m>m_*$ where one can see the changes in the
number of event horizons according to $R$ and ${\cal C}$,
respectively.
\begingroup
\setlength{\tabcolsep}{6pt} 
\renewcommand{\arraystretch}{1.8}
\begin{table}[ht!]
  \begin{center}
    \caption{Numbers of event horizons of the HRBH in holographic massive gravity according to $R$ and
${\cal C}$ when $m>m_*$. Note that in the table, max(2) or max(3)
denote the cases that maximum two or three horizons are formed
satisfying $f(r_H)=0$ in the given range, respectively.}
  \label{table1}
    \begin{tabular}{|c|c|c|c|c|c|}
    \hline
           &  ${\cal C}<-1$ & ${\cal C}=-1$ &  $-1<{\cal C}<0$  &  ${\cal C}=0$ &  ${\cal C}>0$ \\
    \hline
      $R>0$ & {\rm max}(3) & {\rm max}(2)    & {\rm max}(2)  & {\rm max}(2)   & {\rm max}(2)  \\
     $R=0$  & 0 & 0    & 2  &  2 & {\rm max}(2)  \\
     $R<0$  & 0 & 0    &  {\rm max}(3)  &   {\rm max}(3) & {\rm max}(3)  \\
    \hline
    \end{tabular}
  \end{center}
\end{table}
\endgroup
\begingroup
\setlength{\tabcolsep}{6pt} 
\renewcommand{\arraystretch}{1.8}
\begin{table}[ht!]
  \begin{center}
    \caption{Numbers of event horizons of the HRBH in holographic massive gravity according to $R$ and
${\cal C}$ when $m=m_*$.}
     \label{table2}
    \begin{tabular}{|c|c|c|c|c|c|}
    \hline
           &  ${\cal C}<-1$ & ${\cal C}=-1$ &  $-1<{\cal C}<0$  &  ${\cal C}=0$ &  ${\cal C}>0$ \\
    \hline
      $R>0$ & 1 & 1    & {\rm max}(2)  &  0 & 0  \\
     $R=0$  & 0 & 0    & 2  &  1 & 0 \\
     $R<0$  & 0 & 0    & {\rm max}(3) &  {\rm max}(3)  &  {\rm max}(3)  \\
    \hline
    \end{tabular}
  \end{center}
\end{table}
\endgroup

One can see from the Tables \ref{table1}, \ref{table2}, and Fig.
\ref{fig1} that when $m>m_*$, it would be physically interesting
in both all ranges of ${\cal C}$ with $R>0$ and all ranges of $R$
with ${\cal C}>-1$. In those ranges, there exists not only an
outer, but also at least one inner event horizon. When $m=m_*$,
the extremal case for the HRBH in massless gravity, it remains
extremal for ${\cal C}\le -1$ and $R>0$ in the HRBH in massive
gravity. However, for all ranges of $R$ with $-1<{\cal C}<0$ and
$R<0$ with ${\cal C}\ge 0$, it changes to have two and more event
horizons. As a result, according to the graviton mass parameters
$R$ and ${\cal C}$, it is expected that the thermodynamics of the
HRBH in massive gravity is differently described from the HRBH in
massless gravity.

\begin{figure*}[t!]
   \centering
   \includegraphics{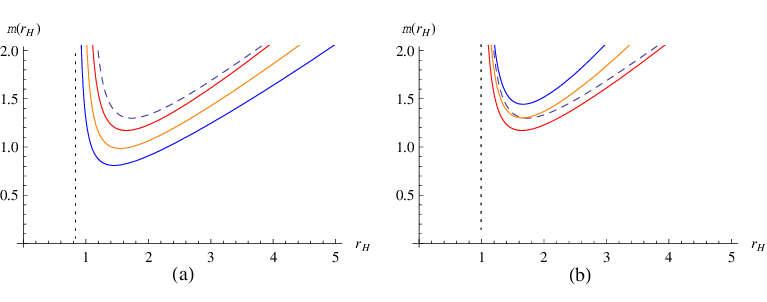}
\caption{Mass function $m(r_H)$: (a) $(R, {\cal
C})=(0,0),~(0.01,-0.1),~ (0.01,-0.2),~ (0.01,-0.3)$ from top to
bottom curves,  (b) $(R, {\cal C})=(0.01,-0.1),~(0,0),~
(0.03,-0.1),~(0.05,-0.1)$ from bottom to top curves. The dashed
lines corresponding to $(R, {\cal C})=(0,0)$ are for the HRBH in
massless gravity.}
 \label{fig2}
\end{figure*}

On the other hand, from the metric solution (\ref{lapsewol}), one
can find that the event horizon $r_H$ determines the mass as
 \be
 m(r_H)=\frac{(1+{\cal C}+2Rr_H)r^3_H}{2[r^2_H-(1+{\cal C}+2Rr_H)l^2]}.
 \ee
In Fig. \ref{fig2}, we have plotted the mass function by comparing
the HRBH in massless gravity with ones in holographic massive
gravity. As a result, one can see that for a fixed $R$ (or $\cal
C$), as $\cal C$ (or $R$) decreases, the outer event horizon $r_H$
increases in the HRBH in massive gravity.

Finally, it is appropriate to comment on the Kretschmann scalar,
which is known to be finite for regular black holes, for the HRBH
in massive gravity. First of all, for the HRBH in massless gravity
\cite{Bambi:2013ufa}, the Kretschmann scalar is given by
  \bea\label{Kret}
 {\cal R}_{\mu\nu\rho\sigma}{\cal R}^{\mu\nu\rho\sigma}
      =\frac{48m^2(r^{12}-8l^2mr^9+72l^4m^2r^6-16l^6m^3r^3+32l^8m^4)}{(r^3+2ml^2)^6}.
 \eea
At the center of the spacetime as $r\rightarrow 0$, due to the
Hayward parameter $l$, it is finite as
 \bea
 {\cal R}_{\mu\nu\rho\sigma}{\cal R}^{\mu\nu\rho\sigma}
      &=&\frac{24}{l^4},
 \eea
which shows that the HRBH in massless gravity is regular
everywhere with no curvature singularity. On the other hand, as
$l\rightarrow 0$, it becomes
 \bea
 {\cal R}_{\mu\nu\rho\sigma}{\cal R}^{\mu\nu\rho\sigma}
      =\frac{48m^2}{r^6},
 \eea
the Kretschmann scalar of the Schwarzschild spacetime in massless
gravity, as expected.

Now, when massive gravitons are introduced as the HRBH in massive
gravity having the lapse function (\ref{lapsewol}), we have new
Kretschmann scalar modified by $R$ and ${\cal C}$ as
 \bea\label{mKret}
 {\cal R}_{\mu\nu\rho\sigma}{\cal R}^{\mu\nu\rho\sigma}
      &=&\frac{48m^2(r^{12}-8l^2mr^9+72l^4m^2r^6-16l^6m^3r^3+32l^8m^4)}{(r^3+2ml^2)^6}\nonumber\\
      &-&\frac{192l^2m^2R}{r(r^3+2ml^2)^2}
         -\frac{16m{\cal C}}{r^2(r^3+2ml^2)}+\frac{32R^2}{r^2}+\frac{16R{\cal C}}{r^3}+\frac{4{\cal C}^2}{r^4}.
 \eea
In the massless limit of $R=0$ and ${\cal C}=0$, it becomes the
Kretschmann scalar (\ref{Kret}) of the HRBH in massless gravity.
However, near the center as $r\rightarrow 0$, the last term in Eq.
(\ref{mKret}) becomes dominant so it diverges as
 \be
 {\cal R}_{\mu\nu\rho\sigma}{\cal R}^{\mu\nu\rho\sigma}\sim\frac{4{\cal C}^2}{r^4}\rightarrow\infty.
 \ee
Thus, the regular structure of the HRBH in massless gravity has
been changed to singular due to massive gravitons, less-severely
divergent than the Schwarzschild case though. Note that in the
massive Schwarzschild limit as $l\rightarrow 0$, the Kretschmann
scalar is reduced to
 \be
  {\cal R}_{\mu\nu\rho\sigma}{\cal R}^{\mu\nu\rho\sigma}
  =\frac{32R^2}{r^2}+\frac{16R{\cal C}}{r^3}+\frac{4{\cal C}^2}{r^4}-\frac{16m{\cal C}}{r^5}+\frac{48m^2}{r^6}.
 \ee

As a result, we have found that introducing massive gravitons to
the HRBH in massless gravity affects the spacetime structure from
regular to singular. This singular structure of spacetime can be
relieved by enlarging the dimensions of the spacetime according to
the well established GEMS scheme, which we will consider in
Section III.

\subsection{Solutions of the HRBH in massless gravity}

\begin{figure*}[t!]
   \centering
   \includegraphics{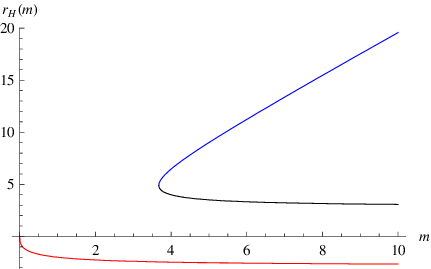}
\caption{Solutions for the HRBH in massless gravity: the blue
curve is for $\tilde{r}_{H2}$, the black curve for
$\tilde{r}_{H3}$ and the red curve for $\tilde{r}_{H1}$ with
$l=1$. Here, we see that $\tilde{r}_{H2}$ corresponds to an outer
horizon ($r_+$) and $\tilde{r}_{H3}$ to an inner horizon ($r_-$).
$\tilde{r}_{H1}$ is negative so discarded. }
 \label{fig3}
\end{figure*}

Let us first briefly summarize the solutions of the HRBH in
massless gravity satisfying $f(\tilde{r}_H)=0$ in Eq.
(\ref{lapsewol}) where $\tilde{r}_H$ denotes the horizon radius of
the black hole in massless gravity. In the massless case with the
graviton mass parameters $R={\cal C}=0$, event horizons can be
found from the following cubic equation rewritten as
 \be\label{Hayward-massless}
 \tilde{r}^3_H-2m\tilde{r}^2_H+2l^2m=0.
 \ee
By following the general procedure from Eq. (\ref{standardcubic})
to Eq. (\ref{standardcubicsol}) presented in the next subsection,
one finally gets the solutions of the HRBH in massless gravity as
 \bea\label{cubic-sol1}
 \tilde{r}_{H1}&=&\frac{2m}{3}\left[1-2\cos\left(\frac{\psi}{3}\right)\right],\nonumber\\
 \tilde{r}_{H2}&=&\frac{2m}{3}\left[1-2\cos\left(\frac{\psi}{3}+\frac{2\pi}{3}\right)\right],\nonumber\\
 \tilde{r}_{H3}&=&\frac{2m}{3}\left[1-2\cos\left(\frac{\psi}{3}+\frac{4\pi}{3}\right)\right],
 \eea
whose behaviors are depicted in Fig. \ref{fig3}. Here, $\cos\psi$
is defined as
 \be\label{cospsi01}
 \cos\psi=-1+\frac{27l^2}{8m^2}.
 \ee
Note that the solutions (\ref{cubic-sol1}) satisfy the properties
 \bea
 \tilde{r}_{H1}+\tilde{r}_{H2}+\tilde{r}_{H3}&=&2m,\nonumber\\
 \tilde{r}_{H1}\tilde{r}_{H2}+\tilde{r}_{H2}\tilde{r}_{H3}+\tilde{r}_{H3}\tilde{r}_{H1}&=& 0,\nonumber\\
 \tilde{r}_{H1}\tilde{r}_{H2}\tilde{r}_{H3} &=&-2l^2m.
 \eea

\subsection{Solution of the HRBH in massive gravity}

In this subsection, we will newly find general solutions of the
HRBH in holographic massive gravity (\ref{lapsewol}) satisfying
$f(r_H)=0$ where $r_H$ denotes the horizon radius of the black
hole in massive gravity. First of all, one can rewrite $f(r_H)=0$
as
 \be\label{quartic}
 2Rr^4_H+(1+{\cal C})r^3_H-2mr^2_H+4Rl^2mr_H+2l^2m(1+{\cal C})=0,
 \ee
which is a quartic equation with $R\neq 0$, compared with the
cubic one in the previous massless case. After dividing by $2R$
and by introducing a new variable $y$ as
 \be\label{quartic-eq-gen}
 y=r_H+\frac{1+{\cal C}}{8R},
 \ee
one can find the standard quartic equation written as
 \be
 \label{quartic-eq}
 y^4+a_1 y^2+a_2y+a_3=0,
 \ee
where
 \bea
 a_1&=&-\frac{m}{R}-\frac{3(1+{\cal C})^2}{32R^2},\nonumber\\
 a_2&=& 2l^2m+\frac{(1+{\cal C})m}{4R^2}+\frac{(1+{\cal C})^3}{64R^3},\nonumber\\
 a_3&=& \frac{3(1+{\cal C})l^2m}{4R}-\frac{(1+{\cal C})^2m}{64R^3}-\frac{3(1+{\cal C})^4}{4096R^4}.
 \eea
Note that at this stage one cannot simply take the limit of
$R\rightarrow 0$ since the coefficients are all divergent. Thus,
if one wants to find the limit, it should consider the equation
(\ref{quartic}) again from the start.

The quartic equation (\ref{quartic-eq}) can be solved by adding
and subtracting $xy^2+x^2/4$ to Eq. (\ref{quartic-eq}) as
 \be
 \left(y^4+xy^2+\frac{1}{4}x^2\right)-xy^2-\frac{1}{4}x^2+a_1 y^2+a_2y+a_3=0,
 \ee
which can be rewritten as
 \bea
 0&=&\left(y^2+\frac{1}{2}x\right)^2-\left[(x-a_1)y^2-a_2y+\left(\frac{1}{4}x^2-a_3\right)\right]\nonumber\\
  &=&\left(y^2+\frac{1}{2}x\right)^2-\left(\sqrt{x-a_1}y-\frac{a_2}{2\sqrt{x-a_1}}\right)^2
  +\frac{a^2_2}{4(x-a_1)}-\left(\frac{1}{4}x^2-a_3\right).
 \eea
Thus, if we demand the last two terms to vanish as
 \be\label{qubic-eq}
 \frac{a^2_2}{4(x-a_1)}-\left(\frac{1}{4}x^2-a_3\right)=0,
 \ee
one can have
 \bea
 0&=&\left(y^2+\frac{1}{2}x\right)^2-\left(\sqrt{x-a_1}y-\frac{a_2}{2\sqrt{x-a_1}}\right)^2\nonumber\\
 &=&\left(y^2-\sqrt{x-a_1}y+\frac{1}{2}x+\frac{a_2}{2\sqrt{x-a_1}}\right)\left(y^2+\sqrt{x-a_1}y+\frac{1}{2}x-\frac{a_2}{2\sqrt{x-a_1}}\right).
 \eea
These are the product of two quadratic equations whose roots can
be easily obtained separately. As a result, we have the following
four roots for the quartic equation (\ref{quartic-eq})
 \bea
 y_1=\frac{1}{2}(p_1+p_2),~~ y_2=\frac{1}{2}(p_1-p_2),~~ y_3&=&\frac{1}{2}(-p_1+p_3),~~ y_4=\frac{1}{2}(-p_1-p_3),
 \eea
with
 \bea\label{sol-ref1}
 p_1\equiv(x_0-a_1)^{1/2},~~p_2\equiv\left(-p^2_1-2a_1-\frac{2a_2}{p_1}\right)^{1/2},~~p_3\equiv\left(-p^2_1-2a_1+\frac{2a_2}{p_1}\right)^{1/2}.
 \eea
In Eq. (\ref{sol-ref1}), note that $x_0$ is a root of the cubic
equation (\ref{qubic-eq}) rewritten as
 \be\label{cubic-eq3}
 x^3-a_1x^2-4a_3x+4a_1a_3-a^2_2=0.
 \ee
Then, we have
 \be
 b_1=-a_1,~~b_2=-4a_3,~~b_3=4a_1a_3-a^2_2,
 \ee
where $b_1,~b_2$, and $b_3$ are the coefficients of the standard
cubic equation of
 \be\label{standardcubic}
  x^3+b_1x^2+b_2x+b_3=0.
 \ee
These give us
  \be
 q_1=\frac{9b_1b_2-27b_3-2b^3_1}{54},~~q_2=\frac{3b_2-b^2_1}{9}
 \ee
to yield
 \bea\label{q12s}
 q_1&=&2l^4m^2+\frac{3(1+{\cal C})l^2m^2}{2R^2}-\frac{m^3}{27R^3}+\frac{(1+{\cal C})^3l^2m}{8R^3},\nonumber\\
 q_2&=&-\frac{(1+{\cal C})l^2m}{R}-\frac{m^2}{9R^2}.
 \eea
Now, we define $\cos\psi$ \cite{Hong:2013bca,Weisstein2003} as
 \be\label{cospsi0}
 \cos\psi=\frac{q_1}{(-q^3_2)^{1/2}}.
 \ee
Making use of $q_1$ and $q_2$ in Eq. (\ref{q12s}), we find
 \bea\label{cospsi}
 \cos\psi=\frac{-8m^3+27(1+{\cal C})^3l^2m+324(1+{\cal C})l^2m^2R+432l^4m^2R^3}{8[m^2+9(1+{\cal
 C})l^2mR]^{3/2}}.
 \eea

Then, the solutions for the cubic equation (\ref{cubic-eq3}) are
given by
 \bea\label{standardcubicsol}
 x_1&=&\frac{2}{3R}[m^2+9(1+{\cal C})l^2mR]^{1/2}\cos\left(\frac{\psi}{3}\right)-\frac{1}{3R}\left[m+\frac{3(1+{\cal C})^2}{32R}\right],\nonumber\\
 x_2&=&\frac{2}{3R}[m^2+9(1+{\cal C})l^2mR]^{1/2}\cos\left(\frac{\psi}{3}+\frac{2\pi}{3}\right)-\frac{1}{3R}\left[m+\frac{3(1+{\cal C})^2}{32R}\right],\nonumber\\
 x_3&=&\frac{2}{3R}[m^2+9(1+{\cal C})l^2mR]^{1/2}\cos\left(\frac{\psi}{3}+\frac{4\pi}{3}\right)-\frac{1}{3R}\left[m+\frac{3(1+{\cal C})^2}{32R}\right].
 \eea
Thus, making use of these solutions, we finally have the following
four roots of the quartic equation (\ref{quartic-eq}) as
 \bea\label{quartic-sols}
 r_{H1}&=&\frac{1}{2}\left(p_1+p_2-\frac{1+{\cal C}}{4R}\right),\nonumber\\
 r_{H2}&=&\frac{1}{2}\left(p_1-p_2-\frac{1+{\cal C}}{4R}\right),\nonumber\\
 r_{H3}&=&\frac{1}{2}\left(-p_1+p_3-\frac{1+{\cal C}}{4R}\right),\nonumber\\
 r_{H4}&=&\frac{1}{2}\left(-p_1-p_3-\frac{1+{\cal C}}{4R}\right),
 \eea
where
 \bea
 p_1&=&\left(\frac{2}{3R}[m^2+9(1+{\cal C})l^2mR]^{1/2}\cos\left(\frac{\psi}{3}\right)+\frac{2}{3R}\left[m+\frac{3(1+{\cal C})^2}{32R}\right]\right)^{1/2},\nonumber\\
 p_2&=&\left(-\frac{2}{3R}[m^2+9(1+{\cal C})l^2mR]^{1/2}\cos\left(\frac{\psi}{3}\right)+\frac{4}{3R}\left[m+\frac{3(1+{\cal C})^2}{32R}\right]\right.\nonumber\\
    &&\left.-2\left\{2l^2m+\frac{(1+{\cal C})m}{4R^2}+\frac{(1+{\cal C})^3}{64R^3}\right\}
    \left\{\frac{2}{3R}[m^2+9(1+{\cal C})l^2mR]^{1/2}\cos\left(\frac{\psi}{3}\right)+\frac{2}{3R}\left[m+\frac{3(1+{\cal C})^2}{32R}\right]\right\}^{-1/2}\right)^{1/2},\nonumber\\
 p_3&=&\left(-\frac{2}{3R}[m^2+9(1+{\cal C})l^2mR]^{1/2}\cos\left(\frac{\psi}{3}\right)+\frac{4}{3R}\left[m+\frac{3(1+{\cal C})^2}{32R}\right]\right.\nonumber\\
    &&\left.+2\left\{2l^2m+\frac{(1+{\cal C})m}{4R^2}+\frac{(1+{\cal C})^3}{64R^3}\right\}
    \left\{\frac{2}{3R}[m^2+9(1+{\cal C})l^2mR]^{1/2}\cos\left(\frac{\psi}{3}\right)+\frac{2}{3R}\left[m+\frac{3(1+{\cal C})^2}{32R}\right]\right\}^{-1/2}\right)^{1/2},\nonumber\\
 \eea
Here, we have chosen $x_1$ as $x_0$ in $p_1$.
\begin{figure*}[t!]
   \centering
   \includegraphics{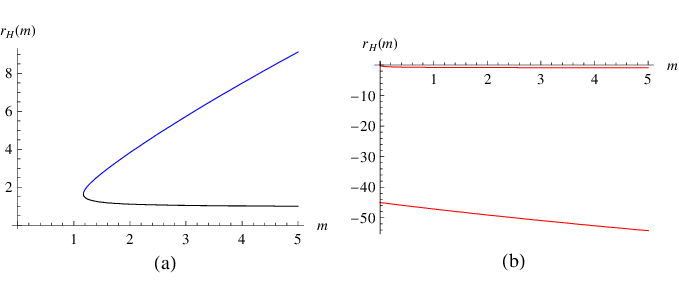}
\caption{Solutions for the HRBH in massive gravity: in (a) the
blue curve is for $r_{H1}$, the black curve for $r_{H2}$, which
correspond to an outer horizon ($r_+$) and to an inner horizon
($r_-$), respectively. In (b), the curves are for $r_{H3}$ (upper)
and $r_{H4}$ (lower), which are negative so discarded. Here, we
set $R=0.01$, ${\cal C}=-0.1$ with $l=1$. }
 \label{fig4}
\end{figure*}
\begin{figure*}[t!]
   \centering
   \includegraphics{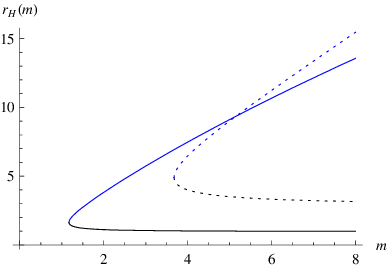}
\caption{Solutions for the HRBH in massless and massive gravity:
the blue and black solid curves are for massive gravity, while the
dotted curves for massless gravity. Here, we set $R=0.01$, ${\cal
C}=-0.1$ with $l=1$.}
 \label{fig5}
\end{figure*}

In Fig. \ref{fig4}, we depict a set of solutions for the HRBH in
massive gravity for $R=0.01$ and ${\cal C}=-0.1$. As explained in
Table \ref{table1}, for the chosen $R$ and ${\cal C}$, we expect
that there are two event horizons, and Fig. \ref{fig4} shows the
same behavior that there are two, i.e., one is an outer $r_{H1}$
and the other is an inner horizon $r_{H2}$, respectively. The
remaining two $r_{H3}$, $r_{H4}$ are of no physical meanings for
event horizons since they are negative. In Fig. \ref{fig5}, we
also draw the solution for the HRBH in massive gravity compared
with the HRBH in massless gravity, where one can see how massive
gravitons change $r_H(m)$.

Note that $r_{H_i}~(i=1,2,3,4)$ in Eq. (\ref{quartic-sols}) denote
the event horizons of the HRBH in holographic massive gravity,
while $\tilde{r}_{H_i}~(i=1,2,3)$ in Eqs. (\ref{cubic-sol1}) stand
for those of the HRBH in massless gravity. It is also appropriate
to comment that it is not possible directly to get the solutions
of the HRBH in massless gravity by taking $R\rightarrow 0$ and
${\cal C}\rightarrow 0$ from the solutions (\ref{quartic-sols}) of
the quartic equation. Note that this can be understood due to the
fact that they are obtained from the implicit condition of $R\neq
0$ as in Eqs. (\ref{quartic}) and (\ref{quartic-eq-gen}).

\section{GEMS embedding of HRBH}
\setcounter{equation}{0}
\renewcommand{\theequation}{\arabic{section}.\arabic{equation}}

\subsection{HRBH in massless gravity}

The (3+1)-dimensional HRBH in massless gravity can be embedded in
a (5+2)-dimensional Minkowski spacetime given by
 \be
 ds^{2}= \eta_{IJ}dz^Idz^J,~{\rm with}~\eta_{IJ}={\rm
 diag}(-1,1,1,1,1,1,-1),
 \ee
where embedding coordinates are obtained as
 \bea\label{gems-Haywardmassless}
 z^{0}&=&\tilde{k}_{H}^{-1}f^{1/2}(r)\sinh \tilde{k}_{H}t, \nonumber \\
 z^{1}&=&\tilde{k}_{H}^{-1}f^{1/2}(r)\cosh \tilde{k}_{H}t, \nonumber \\
 z^{2}&=&r\sin\theta\cos\phi, \nonumber \\
 z^{3}&=&r\sin\theta\sin\phi, \nonumber \\
 z^{4}&=&r\cos\theta, \nonumber \\
 z^{5}&=&\int    \frac{dr}{2\tilde{k}_H}
                 \left(\frac{H_0[r^8\tilde{r}^{13}_H+l^4r^2\tilde{r}^9_H(30r^6+5r^3\tilde{r}^3_H+4\tilde{r}^6_H)+20l^6r^5\tilde{r}^{10}_H+l^8r^2\tilde{r}^5_H(33r^6+6\tilde{r}^6_H)+18l^{10}r^5\tilde{r}^6_H]}
                            {\tilde{r}^6_H(r^2\tilde{r}^2_H-l^2H_0)[r^3(\tilde{r}^2_H-l^2)+l^2\tilde{r}^3_H]^3}\right)^{1/2},\nonumber\\
z^{6}&=& \int\frac{ldr}{2\tilde{k}_H}
                 \left(\frac{H_0[r^5\tilde{r}^{11}_H(9r^3+4\tilde{r}^3_H)+l^4r^2\tilde{r}^7_H(46r^6+\tilde{r}^6_H)+39l^6r^5\tilde{r}^8_H+9l^8r^2\tilde{r}^3_H(r^6+\tilde{r}^6_H)]}
                            {\tilde{r}^6_H(r^2\tilde{r}^2_H-l^2H_0)[r^3(\tilde{r}^2_H-l^2)+l^2\tilde{r}^3_H]^3}\right)^{1/2},
 \eea
with
 \bea\label{H0}
  H_0= r^2+r\tilde{r}_H+\tilde{r}^2_H.
 \eea
In the above embedding functions (\ref{gems-Haywardmassless}),
$\tilde{k}_H$ is the surface gravity defined as
 \be\label{sgravity}
 \tilde{k}_H = \left.\sqrt{-\frac{1}{2}(\na^{\mu}\xi^{\nu})(\na_{\mu}\xi_{\nu})}~\right|_{r= \tilde{r}_H}
     =\frac{\tilde{r}^2_H-3l^2}{2\tilde{r}^3_H},
 \ee
where $\xi^\mu$ is a Killing vector and $\tilde{r}_H$ is the event
horizon of the HRBH in massless gravity. Note that in the limit of
$l\rightarrow 0$, $z^6$ vanishes, and $z^5$ becomes
 \be
  z^{5}=\int \frac{dr}{2\tilde{k}_H}
              \left(\frac{r^2+r\tilde{r}_H+\tilde{r}^2_H}{\tilde{r}_Hr^3}\right)^{1/2}.
 \ee
Therefore, with $z^0~\cdot\cdot\cdot~z^4$ expressed in the same
limit, the embedding coordinates (\ref{gems-Haywardmassless}) are
correctly reduced to the well-known (5+1)-dimensional GEMS
embeddings of the Schwarzschild black hole
\cite{Fronsdal:1959zza}.

Here, we note that from the original spacetime metric, the Hawking
temperature $\tilde{T}_H$ seen by an asymptotic observer and  a
local fiducial temperature measured by an observer who rests at a
distance from the black hole are simply found as
 \be\label{HawkingT0}
 \tilde{T}_H=\frac{\tilde{k}_H}{2\pi}=\frac{\tilde{r}^2_H-3l^2}{4\pi\tilde{r}^3_H},
 \ee
 \be\label{fidT0}
 \tilde{T}_{\rm FID}(r)=\frac{\tilde{T}_H}{\sqrt{f(r)}}=\frac{(\tilde{r}^2_H-3l^2)[r^3(\tilde{r}^2_H-l^2)+l^2\tilde{r}^3_H]^{1/2}}
                {4\pi\tilde{r}^3_H[(r-\tilde{r}_H)(r^2\tilde{r}^2_H-H_0l^2)]^{1/2}},
 \ee
respectively.

Now, let us consider the Unruh effect in the embedded flat
spacetime, which originally states that accelerated observers or
detectors with an acceleration $a$ along the $x$ direction by
following the trajectory $a^{-2}=x^2-t^2$ measure the Unruh
temperature given by $2\pi T=a$ \cite{Unruh:1976db}. To apply this
for in a higher dimensional flat spacetime, we notice that the
static detectors ($r,~\theta,~\phi={\rm constant}$) in the
original curved spacetime are described by a fixed point in the
($z^2,~z^3,~z^4,~z^5,~z^6$) plane on the GEMS embedded spacetime.
Then, an observer who is uniformly accelerated in the
(5+2)-dimensional flat spacetime, follows a hyperbolic trajectory
described by
 \be\label{acc7}
 a^{-2}_7=(z^1)^2-(z^0)^2= \frac{f(r)}{\tilde{k}^2_H}.
 \ee
Thus, one can find the Unruh temperature for the uniformly
accelerated observer in the (5+2)-dimensional flat spacetime as
 \be\label{unruh7}
 \tilde{T}_U=\frac{a_7}{2\pi}
    =\frac{(\tilde{r}^2_H-3l^2)[r^3(\tilde{r}^2_H-l^2)+l^2\tilde{r}^3_H]^{1/2}}
                {4\pi \tilde{r}^3_H[(r-\tilde{r}_H)(r^2\tilde{r}^2_H-H_0l^2)]^{1/2}}.
 \ee
This corresponds to the fiducial temperature (\ref{fidT0}) for the
observer located at a distance from the HRBH in massless gravity.
The Hawking temperature $\tilde{T}_H$ seen by an asymptotic
observer can be obtained as
 \be
 \tilde{T}_H=\sqrt{-g_{00}}\tilde{T}_U=\frac{\tilde{k}_H}{2\pi}.
 \ee
As a result, one can see that the Hawking effect for a fiducial
observer in a black hole spacetime is equal to the Unruh effect
for a uniformly accelerated observer in a higher-dimensional flat
spacetime.


Now, let us find a freely falling acceleration and corresponding
temperature in the (5+2)-dimensional embedded flat spacetime. For
an observer who is freely falling from rest $r=r_{0}$ at $\tau=0$,
the equations of motion are
 \bea\label{eomr0}
 \frac{dt}{d\tau}&=&\frac{f^{1/2}(r_0)}{f(r)}=\left(1-\frac{2mr^2_0}{r^3_0+2l^2m}\right)^{1/2}
                                            \left(1-\frac{2mr^2}{r^3+2l^2m}\right)^{-1},\nonumber\\
 \frac{dr}{d\tau}&=&-[f(r_0)-f(r)]^{1/2}=-\left[\frac{-2m\{r^2_0r^2(r-r_0)-2l^2m(r^2-r^2_0)\}}{(r^3_0+2l^2m)(r^3+2l^2m)}\right]^{1/2},
 \eea
where $(-)$ sign is for inward motion. Then, making use of the
embedding coordinates in Eq. (\ref{gems-Haywardmassless}) and the
geodesic equations in Eq. (\ref{eomr0}), one can explicitly find a
freely falling acceleration $\bar{a}_{7}$ in the GEMS embedded
(5+2)-dimensional spacetime as
 \bea\label{a7Haywardmassless}
 \bar{a}^2_{7}&=&\sum_{I=0}^{6}\left.\eta_{IJ}\frac{dz^I}{d\tau}\frac{dz^J}{d\tau}\right|_{r=r_0}\nonumber\\
              &=&\frac{N_1N_2}{4\tilde{r}^6_H(r^2\tilde{r}^2_H-l^2H_0)[r^3(\tilde{r}^2_H-l^2)+l^2\tilde{r}^3_H]^3},
 \eea
where
 \bea\label{d1d2}
 N_1&=&r^4\tilde{r}^6_H(r+\tilde{r}_H)-l^2r\tilde{r}^4_HH_1+l^4\tilde{r}^2_HH_0H_2-3l^6(r-\tilde{r}_H)H^2_0,\nonumber\\
 N_2&=&r^4\tilde{r}^6_H(r^2+\tilde{r}^2_H)-l^2r\tilde{r}^4_HH_3+l^4\tilde{r}^2_H(r-\tilde{r}_H)H_0H_2-3l^6(r-\tilde{r}_H)^2H^2_0
 \eea
with
 \bea
 H_1 &=& 5r^4+5r^3\tilde{r}_H+4r^2\tilde{r}^2_H+2r\tilde{r}^3_H+2\tilde{r}^4_H,\nonumber\\
 H_2 &=& 7r^3-\tilde{r}^3_H, \nonumber\\
 H_3 &=& 5r^5+r^3\tilde{r}^2_H-2r^2\tilde{r}^3_H+2\tilde{r}^5_H.
 \eea
Note here that $r_0$ is replaced with $r$ in Eq.
(\ref{a7Haywardmassless}).

According to the Unruh's prescription, the freely falling
acceleration gives us the freely falling temperature measured by
the freely falling observer as
 \bea \label{tffar-Haywardmassless}
 \tilde{T}_{\rm FF}  &=& \frac{\bar{a}_7}{2\pi}\nonumber\\
                     &=& \frac{1}{4\pi\tilde{r}^3_H}{\sqrt{\frac{N_1N_2}{(r^2\tilde{r}^2_H-l^2H_0)[r^3(\tilde{r}^2_H-l^2)+l^2\tilde{r}^3_H]^3}}}.
 \eea
Making use of the dimensionless parameters $x=\tilde{r}_H/r$ and
$b=l/\tilde{r}_H$, the squared freely falling temperature can be
written as
 \be
 \tilde{T}^2_{\rm FF}=\frac{[1+x-b^2h_1+b^4h_0h_2-3b^6(1-x)h^2_0]
                    [1+x^2-b^2h_3+b^4(1-x)h_0h_2-3b^6(1-x^3)^2]}
              {16\pi^2\tilde{r}^2_H(1-b^2h_0)[1-b^2(1-x^3)]^3},
 \ee
where
 \bea
 h_0 &\equiv& H_0/r^2=1+x+x^2,\nonumber\\
 h_1 &\equiv& H_1/r^4=5+5x+4x^2+2x^3+2x^4,\nonumber\\
 h_2 &\equiv& H_2/r^3=7-x^3, \nonumber\\
 h_3 &\equiv& H_3/r^5=5+x^2-2x^3+2x^5.
 \eea
\begin{figure*}[t!]
   \centering
   \includegraphics{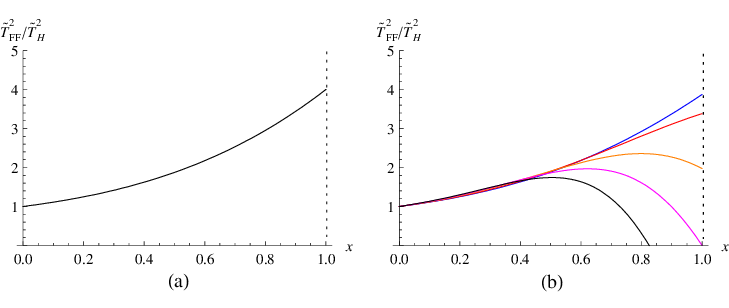}
\caption{Squared freely falling temperatures $\tilde{T}^2_{\rm
FF}/\tilde{T}^2_{\rm H}$ for the HRBH in massless gravity drawn by
a dimensionless parameter $x~(=\tilde{r}_H/r)$. (a) the freely
falling temperature of the Schwarzschild black hole in massless
gravity which corresponds to $b=0$ (or, $l=0$). (b) the freely
falling temperatures of the HRBH in massless gravity for
$b=0.1,~0.2,~0.3,~b_c,~0.4$ from top to bottom. Here, the vertical
dotted lines are drawn at event horizons.}
 \label{fig6}
\end{figure*}
As $r\rightarrow\infty$, the freely falling temperature
$\tilde{T}_{\rm FF}$ is reduced to the Hawking temperature
(\ref{HawkingT0}). Moreover, as $l\rightarrow 0$, it becomes the
freely falling temperature of the Schwarzschild black hole as
 \be
 T^{\rm Sch}_{\rm FF}=\frac{1}{4\pi\bar{r}_H}\sqrt{\frac{r^3+\bar{r}_HH_0}{r^3}},
 \ee
where $\bar{r}_H=2m$ is the radius of the event horizon of the
Schwarzschild black hole. Note also that as $r\rightarrow\infty$,
$T^{\rm Sch}_{\rm FF}$ becomes the Hawking temperature of the
Schwarzschild black hole, while as $r\rightarrow\bar{r}_H$, the
freely falling temperature becomes
 \be
 T^{\rm Sch}_{\rm FF}\rightarrow\frac{1}{2\pi\bar{r}_H},
 \ee
not diverge, but remains finite at the event horizon.

In Fig. \ref{fig6}, we have depicted the ratio of the squared
freely falling temperatures to the squared Hawking temperature,
$\tilde{T}^2_{\rm FF}/\tilde{T}^2_H$. For comparison purposes, it
is shown in Fig. \ref{fig6}(a) that the freely falling temperature
of the Schwarzschild black hole, which is a prototype of a
singular black hole, is finite at the event horizon of $x=1$
($r=\tilde{r}_H$), while it becomes the Hawking temperature at
asymptotic infinity of $x\rightarrow 0$ ($r\rightarrow\infty$). On
the other hand, as in Fig. \ref{fig6}(b), for the HRBH in massless
gravity, the freely falling temperatures are finite only when
$0<b<b_c(\equiv 0.3568)$ at the event horizons, and when $b=b_c$,
it vanishes at the event horizon. Meanwhile, the freely falling
temperatures become the Hawking temperature at asymptotic
infinity. The parameter $b$ also has an upper bound coming from
the Hawking temperature (\ref{HawkingT0}), which is defined when
$\tilde{r}_H\ge\sqrt{3}l$ (or $b\le 1/\sqrt{3}=0.5774$) where
$\tilde{T}^2_{\rm FF}/\tilde{T}^2_H$ diverges. In between
$b_c<b<0.5774$, one can see in Fig. \ref{fig6}(b) that the freely
falling temperatures of the HRBH behave quite differently from the
Schwarzschild singular black hole. They are going up and down, and
then become negative. It is well known that the negativity of
squared freely falling temperatures is not entirely prohibited,
which means that there is no thermal radiation. This is allowed
for a geodesic observer who follows a spacelike motion similar to
the case of the Schwarzschild-AdS black hole in massless gravity
\cite{Deser:1997ri,Brynjolfsson:2008uc,Hong:2019zsi}.

\subsection{HRBH in massive gravity}

Now, after a lengthy calculation, we newly find that the
(3+1)-dimensional HRBH in massive gravity can be embedded in a
(6+3)-dimensional Minkowski spacetime as
 \be
 ds^{2}= \eta_{IJ}dz^Idz^J,~{\rm with}~\eta_{IJ}={\rm
 diag}(-1,1,1,1,1,1,-1,-1,1),
 \ee
whose embedding coordinates are explicitly written as
 \bea\label{gems-Haywardmassive}
 z^{0}&=&k_{H}^{-1}f^{1/2}(r)\sinh k_{H}t, \nonumber \\
 z^{1}&=&k_{H}^{-1}f^{1/2}(r)\cosh k_{H}t, \nonumber \\
 z^{2}&=&r\sin\theta\cos\phi, \nonumber \\
 z^{3}&=&r\sin\theta\sin\phi, \nonumber \\
 z^{4}&=&r\cos\theta, \nonumber \\
 z^{5}&=&\!\!\!\int\!\!\!\frac{dr}{2k_H}\!\left[\frac{r^8r^{12}_HH_5u_1\!+\!l^2r^9r^{10}_HH_5u_2\!+\!l^4r^2r^8_HH^2_5u_3\!+\!2l^6r^4r^6_HH^3_5u_4
                           \!+\!l^8r^5_HH^4_5u_5+3l^{10}r^3r_HH^6_5u_6\!+\!9l^{12}r_HH^8_5u_7}
                 {r^6_H[r^2r^2_H(H_5+2Rr)-l^2H_4H_5H_6][r^3(r^2_H-l^2H_5)+l^2r^3_HH_5]^3}\right]^{1/2},\nonumber\\
  z^{6}&=&\!\!\!\int\!\!\!\frac{dr}{2k_H}\!\left[\frac{r^8r^{12}_HH_5\bar{u}_1\!+\!l^2r^9r^{10}_HH_5\bar{u}_2\!+\!l^4r^2r^8_HH^2_5\bar{u}_3\!+\!2l^6r^4r^6_HH^3_5\bar{u}_4
                           \!+\!l^8r^5_HH^4_5\bar{u}_5+3l^{10}r^3r_HH^6_5\bar{u}_6\!+\!9l^{12}r_HH^8_5\bar{u}_7}
                 {r^6_H[r^2r^2_H(H_5+2Rr)-l^2H_4H_5H_6][r^3(r^2_H-l^2H_5)+l^2r^3_HH_5]^3}\right]^{1/2},\nonumber\\
  z^{7}&=&\!\!\!\int\!\!\!\frac{ldr}{2k_H}\!\left[\frac{r^5r^{11}_HH_5v_1\!+\!6l^2r^4r^{11}_HRH^2_5v_2\!+\!l^4rr^7_HH^3_5v_3\!+\!l^6r^3r^4_HH^4_5v_4
                         \!+\! 3l^8r^2_HH^6_5v_5+9l^{10}H^8_5v_6}
                 {r^6_H[r^2r^2_H(H_5+2Rr)-l^2H_4H_5H_6][r^3(r^2_H-l^2H_5)+l^2r^3_HH_5]^3}\right]^{1/2},\nonumber\\
  z^{8}&=&\!\!\!\int\!\!\!\frac{ldr}{2k_H}\!\left[\frac{r^5r^{11}_HH_5\bar{v}_1\!+\!6l^2r^4r^{11}_HRH^2_5\bar{v}_2\!+\!l^4rr^7_HH^3_5\bar{v}_3\!+\!l^6r^3r^4_HH^4_5\bar{v}_4
                         \!+\! 3l^8r^2_HH^6_5\bar{v}_5+9l^{10}H^8_5\bar{v}_6}
                 {r^6_H[(r^2r^2_H(H_5+2Rr)-l^2H_4H_5H_6][r^3(r^2_H-l^2H_5)+l^2r^3_HH_5]^3}\right]^{1/2}.
 \eea
Here, the surface gravity of the HRBH in massive gravity is given
by
 \be
  k_H = \frac{H_5(r^2_H-3l^2H_5)}{2r^3_H}+R.
 \ee
Also, $H_4$, $H_5$ and $H_6$ are defined as
 \bea
 H_4 &=& 1+{\cal C}+2Rr,\nonumber\\
 H_5 &=& 1+{\cal C}+2Rr_H,\nonumber\\
 H_6&=& r^2+rr_H+r^2_H,
 \eea
and $u_i(\bar{u}_i)~(i=1,2,\cdot\cdot\cdot,7)$ and
$v_i(\bar{v}_i)~(i=1,2,\cdot\cdot\cdot,6)$ are given in Appendix
A. Note that the above GEMS embedding is explicitly carried out
under the assumption of $R\ge 0$ and ${\cal C}\ge -1$, which is
physically much more interesting than others as shown in Figs.
\ref{fig1} and \ref{fig2}, and Tables \ref{table1} and
\ref{table2}.

It seems appropriate to comment that in the massless gravity limit
of $R\rightarrow 0$ and ${\cal C}\rightarrow 0$, $H_4$ and $H_5$
become unity, and $H_6$ becomes $H_0$ in Eq. (\ref{H0}). Also,
when subtracting the embedding coordinates $z^6$ from $z^5$, the
coefficients of $u_i-\bar{u}_i~(i=1,2,\cdot\cdot\cdot,7)$ in the
numerator of $z^5-z^6$  are reduced to Eq. (\ref{diffu}) in the
Appendix. In the same way, by subtracting the embedding
coordinates $z^8$ from $z^7$, the coefficients of
$v_i-\bar{v}_i~(i=1,2,\cdot\cdot\cdot,6)$ in the numerator of
$z^7-z^8$ becomes  Eq. (\ref{diffv}). Since $H_6$ is reduced to
$H_0$ in the massless limit, one can find that they are exactly
the same coefficients in $z^5$ and $z^6$, respectively, of the
HRBH in massless gravity in Eqs. (\ref{gems-Haywardmassless}). As
a result, one can see that when $R\rightarrow 0$ and ${\cal
C}\rightarrow 0$, the (6+3)-dimensional embedding coordinates
(\ref{gems-Haywardmassive}) of the HRBH in massive gravity have
proper limits of the (5+2)-dimensional ones
(\ref{gems-Haywardmassless}) in massless gravity. Moreover, when
one additionally takes the limit of $l\rightarrow 0$, $z^7$ and
$z^8$ identically vanish, and the combination of $z^5$ and $z^6$
by subtraction is finally reduced to
 \be
 z^5=\int
 dr\sqrt{\frac{\bar{r}_H(r^2+r\bar{r}_H+\bar{r}^2_H)}{r^3}},
 \ee
where $\bar{r}_H=2m$ is the event horizon of the Schwarzschild
black hole. Thus, one can see that the embedding coordinate of
$z^5-z^6$ is finally reduced to $z^5$, which is one of the
spacelike embedding coordinates of the Schwarzschild black hole in
the (5+1)-dimensional GEMS scheme.

Here, we note again that from the original spacetime metric, the
Hawking temperature $T_H$ seen by an asymptotic observer can be
found as
 \be\label{HawkingT}
 T_H= \frac{H_5(r^2_H-3l^2H_5)+2 R r^3_H}{4\pi r^3_H},
 \ee
and a local fiducial temperature measured by an observer who rests
at a distance from the black hole is given by
 \be
 T_{\rm FID}(r)=\frac{T_H}{\sqrt{f(r)}}
               =\frac{H_5(r^2_H-3l^2H_5)[r^3(r^2_H-H_5l^2)+l^2r^3_HH_5]^{1/2}}
                {4\pi r^3_H[H_5(r-r_H)(r^2r^2_H-l^2H_6)]^{1/2}}.
 \ee
In the massless limit of $R\rightarrow 0$ and ${\cal C}\rightarrow
0$, one can easily find that it becomes the fiducial temperature
(\ref{fidT0}) since $H_5\rightarrow 1$ and $H_6\rightarrow H_0$.

On the other hand, in order to investigate the Unruh effect in the
GEMS embedded flat spacetime, we note that the static detectors
($r,~\theta,~\phi={\rm constant}$) in the original curved
spacetime are described by a fixed point in the
($z^2,~z^3,~z^4,~z^5,~z^6,~z^7,~z^8$) plane on the GEMS embedded
spacetime. Then, an observer who is uniformly accelerated in the
(6+3)-dimensional flat spacetime, follows a hyperbolic trajectory
in ($z^0,~z^1$) described by
 \be\label{acc9}
 a^{-2}_9=(z^1)^2-(z^0)^2= \frac{f(r)}{k^2_H}.
 \ee
Thus, as before, one can arrive at the Unruh temperature for the
uniformly accelerated observer in the (6+3)-dimensional flat
spacetime as
 \bea\label{unruh9}
 T_U = \frac{a_9}{2\pi}
     = \frac{H_5(r^2_H-3l^2H_5)[r^3(r^2_H-H_5l^2)+l^2r^3_HH_5]^{1/2}}
                {4\pi r^3_H[H_5(r-r_H)(r^2r^2_H-l^2H_6)]^{1/2}}.
  \eea
This corresponds to the fiducial temperature for the observer
located at a distance from the HRBH in massive gravity. The
Hawking temperature $T_H$ seen by an asymptotic observer can be
obtained as
 \be
 T_H=\sqrt{-g_{00}}T_U=\frac{k_H}{2\pi}.
 \ee
As a result, one can see that the Hawking effect for a fiducial
observer in a black hole spacetime is equal to the Unruh effect
for a uniformly accelerated observer in a higher-dimensional flat
spacetime.

Now, let us find a freely falling acceleration and corresponding
temperature in the (6+3)-dimensional embedded flat spacetime. For
an observer who is freely falling from rest $r=r_{0}$ at $\tau=0$,
the equations of motion are
 \bea\label{eomr1}
 \frac{dt}{d\tau}=\frac{f^{1/2}(r_0)}{f(r)}=\left(1-\frac{2mr^2_0}{r^3_0+2l^2m}+2Rr_0+{\cal C}\right)^{1/2}
                                            \left(1-\frac{2mr^2}{r^3+2l^2m}+2Rr+{\cal C}\right)^{-1},\\
 \frac{dr}{d\tau}=-[f(r_0)-f(r)]^{1/2}=-\left[\frac{-2m\{r^2_0r^2(r-r_0)-2l^2m(r^2-r^2_0)\}}{(r^3_0+2l^2m)(r^3+2l^2m)}-2R(r-r_0)\right]^{1/2}.
 \eea
Then, making use of the embedding coordinates in Eq.
(\ref{gems-Haywardmassive}) and the geodesic equations in Eq.
(\ref{eomr1}), one can explicitly find a freely falling
acceleration $a_{9}$ in the GEMS embedded (6+3)-dimensional
spacetime as
 \bea\label{a7Haywardmassive}
 \bar{a}^2_{9}&=&\sum_{I=0}^{8}\left.\eta_{IJ}\frac{dz^I}{d\tau}\frac{dz^J}{d\tau}\right|_{r=r_0}\nonumber\\
              &=&\frac{H_5N_3N_4}{4r^6_H(r^2r^2_H(H_5+2Rr)-l^2H_4H_5H_6)[r^3(r^2_H-l^2H_5)+l^2r^3_HH_5]^3},
 \eea
where
 \bea\label{a7Haywardmassive-numerator}
 N_3 &\equiv& r^4r^6_H(r+r_H)-l^2rr^4_HH_1H_5+l^4r^2_HH_2H^2_5H_6-3l^6(r-r_H)H^3_5H^2_6,\nonumber\\
 N_4 &\equiv& r^4r^6_H[(r^2+r^2_H)H_5+4r^2Rr_H]\nonumber\\
     &-&l^2rr^4_H[(5H_5+8Rr_H)r^5+H_5r^3r^2_H-2(H_5+4Rr_H)r^2r^3_H+2H_5r^5_H]H_5 \nonumber\\
     &+&l^4r^2_H(r-r_H)[6H_5r^3+(r^3-r^3_H)(1+{\cal C}+6Rr_H)]H^2_5H_6  \nonumber\\
     &-&3l^6(r-r_H)^2H^4_5H^2_6.
 \eea
One can easily check that in the massless limit of $R\rightarrow
0$ and ${\cal C}\rightarrow 0$, $D_3$ and $D_4$ are reduced to
$D_1$ and $D_2$, respectively in Eq. (\ref{d1d2}) and thus the
freely falling acceleration $\bar{a}^2_9$ becomes $\bar{a}^2_7$ in
Eq. (\ref{a7Haywardmassless}) of the HRBH in massless gravity.

According to the Unruh's prescription, one can find $T_{\rm FF}$
measured by the freely falling observer from the freely falling
acceleration as
 \be \label{tffar-Haywardmassive}
 T_{\rm FF}=\frac{\bar{a}_9}{2\pi}
           =\frac{1}{4\pi r^3_H}\sqrt{\frac{H_5N_3N_4}{(r^2r^2_H(H_5+2Rr)-l^2H_4H_5H_6)[r^3(r^2_H-l^2H_5)+l^2r^3_HH_5]^3}}.
 \ee
In the massless limit of $R\rightarrow 0$ and ${\cal C}\rightarrow
0$, this is exactly the same with the previous one, the freely
falling temperature of the HRBH in massless gravity in Eq.
(\ref{tffar-Haywardmassless}). Moreover, as $r\rightarrow\infty$,
the freely falling temperature $T_{\rm FF}$ is reduced to the
Hawking temperature.

\begin{figure*}[t!]
   \centering
   \includegraphics{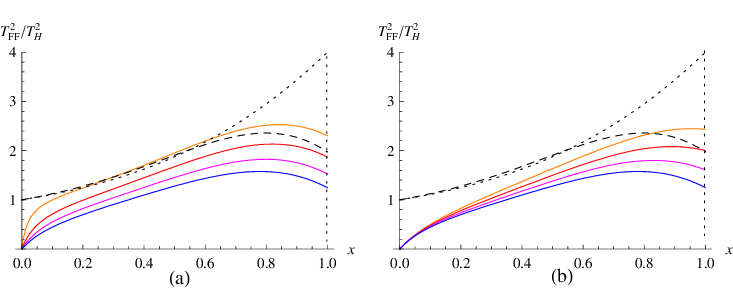}
\caption{Squared freely falling temperatures $T^2_{\rm
FF}/T^2_{\rm H}$ for the HRBH in massive gravity drawn by a
dimensionless parameter $x~(=r_H/r)$. (a) the freely falling
temperature of the HRBH in massive gravity for
$d=0.01,~0.03,~0.05,~0.07$ from top to bottom with a fixed ${\cal
C}=-0.1$ and $b=0.3$. (b) the freely falling temperature of the
HRBH in massive gravity for ${\cal C}=-0.4,-0.3,-0.2,-0.1$ from
top to bottom with a fixed $d=0.07$ and $b=0.3$. Here, the dashed
black curve is for the HRBH in massless gravity as in Fig.
\ref{fig6} with $b=0.3$, and the dotted black curve is for the
Schwarzschild black hole in massless gravity. Also, the vertical
dotted lines are drawn at event horizons.}
 \label{fig7}
\end{figure*}

Now, making use of the dimensionless parameters $x=r_H/r$,
$b=l/r_H$ and $d=Rr_H$, the squared freely falling temperature can
be written as
 \be
 T^2_{\rm FF}=\frac{xn_3n_4}
              {16\pi^2r^2_H(2d+h_5x-b^2h_4h_5h_6x)[1-b^2(1-x^3)h_5]^3},
 \ee
where
 \bea
 n_3&=&1+x-b^2h_1h_5+b^4h_2h^2_5h_6-3b^6(1-x)h^3_5h^2_6,\nonumber\\
 n_4&=&4d+(1+x^2)h_5-b^2[h_3h_5+8d(1-x^3)]h_5+b^4(1-x)[6h_5+(1+{\cal C}+6d)(1-x^3)]h^2_5h_6-3b^6(1-x^3)^2h^4_5\nonumber\\
 \eea
with
 \bea
 h_4&=&1+{\cal C}+\frac{2d}{x},\nonumber\\
 h_5&=&1+{\cal C}+2d,\nonumber\\
 h_6&=&1+x+x^2.
 \eea

In Fig. \ref{fig7}, we have depicted the ratio of the squared
freely falling temperatures to the squared Hawking temperature,
$T^2_{\rm FF}/T^2_H$ for the HRBH in massive gravity. One can see
that at the event horizon the freely falling temperatures are all
finite, while the fiducial temperature diverges
\cite{Brynjolfsson:2008uc}. In the limit of $b\rightarrow 0$ (or
$l\rightarrow 0$), which corresponds to the case of the
Schwarzschild black hole in massive gravity, the freely falling
temperature is reduced to
 \be
 T^2_{\rm FF}=\frac{x(1+{\cal C}+2d)[(1+{\cal C}+2d)(1+x+x^2+x^3)+4d(1+x)]}{16\pi^2r^2_H[(1+{\cal C}+2d)x+2d]}.
 \ee
This is exactly the same with $T^2_{\rm FF}$ in Ref.
\cite{Hong:2019zsi}. Furthermore, in the massless limit of
$d\rightarrow 0$ (or $R\rightarrow 0$) and ${\cal C}\rightarrow
0$, it becomes
 \be
  T^2_{\rm FF}=\frac{1+x+x^2+x^3}{16\pi^2r^2_H},
 \ee
the freely falling temperature of the Schwarzschild black hole in
massless gravity \cite{Brynjolfsson:2008uc}.

\section{Discussion}

In this paper, we have newly studied the Hayward regular black
hole (HRBH) in massive gravity, which is a modification of the
HRBH in massless gravity to have nonzero mass of gravitons as
proposed by Vegh in the framework of holography. By solving
Einstein's equations, we have found a solution of the HRBH in
massive gravity and analyzed the novel structures of event
horizons classified by the graviton mass parameters $R$ and ${\cal
C}$, qualitatively. Concretely, when $m>m_*$, physically
interesting ranges of $R$ and ${\cal C}$ lie in both all ${\cal
C}$ with $R>0$ and all $R$ with ${\cal C}>-1$. In those ranges,
there exist an outer event horizon and at least one inner event
horizon. When $m=m_*$, which is the extremal case for the HRBH in
massless gravity, it remains extremal for ${\cal C}\le -1$ and
$R>0$. However, for all $R$ with $-1<{\cal C}<0$ and $R<0$ with
${\cal C}\ge 0$  in the HRBH in massive gravity, it changes to
have two and more event horizons. Therefore, according to the
specific values in the graviton mass parameters $R$ and ${\cal
C}$, we expect that the thermodynamics of the HRBH in massive
gravity would be different from the one of the HRBH in massless
gravity. Furthermore, to find full information on event horizons,
we have also analytically solved the metric function $f(r_H)=0$ of
the HRBH in massive gravity, which is a non-trivial quartic
equation due to massive gravitons, while the HRBH in massless
gravity is the cubic equation.

After exploiting the geometric property of the spacetime, we have
proceeded to study the GEMS embeddings of the HRBHs in massless
and massive gravity where the former is geodesically incomplete
and the latter is singular due to massive gravitons. As a result,
we have globally embedded the (3+1)-dimensional HRBH in massless
gravity into a (5+2)-dimensional flat Minkowski spacetime, and the
(3+1)-dimensional HRBH in massive gravity into a much higher
(6+3)-dimensional flat Minkowski spacetime, respectively, where
the difference in embedding dimensions comes from whether or not
holographically introduced massive gravitons exist. Furthermore,
making use of newly obtained embedding coordinates, we have
directly obtained Unruh temperatures and compared them with the
Hawking and local fiducial temperatures, showing that the Unruh
effect for a uniformly accelerated observer in a
higher-dimensionally embedded flat spacetime is equal to the
Hawking effect for a fiducial observer in the corresponding curved
spacetime. We have also obtained freely falling temperatures of
the HRBH in massless and massive gravities seen by freely falling
observers following their geodesic trajectories, which remain
finite even at the event horizons. These are different from the
Hawking temperatures divergent at the event horizons.

Finally, it seems appropriate to comment that the solutions of the
HRBH in massless gravity is not possibly obtained simply by taking
$R\rightarrow 0$ and ${\cal C}\rightarrow 0$ due to their implicit
condition of $R\neq 0$ as seen in Eqs. (\ref{quartic}) and
(\ref{quartic-eq-gen}). On the other hand, the GEMS embedding of
the HRBH in massive gravity can be reduced to the corresponding
GEMS embedding of the HRBH in massless gravity in the limit of
$R\rightarrow 0$ and ${\cal C}\rightarrow 0$ by redefining some
embedding coordinates. As a result, when $R\rightarrow 0$ and
${\cal C}\rightarrow 0$, the (6+3)-dimensional embedding
coordinates (\ref{gems-Haywardmassive}) of the HRBH in massive
gravity have proper limits of the (5+2)-dimensional ones
(\ref{gems-Haywardmassless}) in massless gravity, which may be a
characteristic of the Hayward nonsingular black hole in massive
gravity.

\acknowledgments{S. T. H. was supported by Basic Science Research
Program through the National Research Foundation of Korea funded
by the Ministry of Education, NRF-2019R1I1A1A01058449. Y. W. K.
was supported by the National Research Foundation of Korea (NRF)
grant funded by the Korea government (MSIT) (No.
2020R1H1A2102242). }

\appendix
\renewcommand{\theequation}{\thesection.\arabic{equation}}
\section{Coefficients of embedding coordinates of $z^i~(i=5,6,7,8)$}

In this Appendix, we list the coefficients used in
$z^i~(i=5,6,7,8)$ as follows
 \begin{flalign}
 u_1&=r^3H_5(1+{\cal C}+6Rr_H)+4r^2r^2_HR+r_HH_5H_6,&&\nonumber\\
 u_2&=4r^3R[5+34Rr_H+56R^2r^2_H+2(1+{\cal C})(5+17Rr_H)+5{\cal C}^2]+2r^2[2Rr_H(6+41Rr_H+56R^2r^2_H)&&\nonumber\\
    &+(1+{\cal C})(5+55Rr_H+124R^2r^2_H)+{\cal C}^2(10+43Rr_H)+5(1+{\cal C}^3)]+r^2_HH_5R[(1+{\cal C})(r+r_H)+2Rrr_H],&&\nonumber\\
 u_3&=3r^9[13(1+{\cal C})^2+40(1+{\cal C})Rr_H+28R^2r^2_H]+3r^8r_H[(1+{\cal C})(10+7{\cal C}+2R(1+{\cal C})(37r+2r_H))&&\nonumber\\
    &+12Rr_H(6+3{\cal C}+4Rr)+4R^2r_H(30r+7r_H)]+3r^7r^2_H[4Rr_H(1+{\cal C}+4Rr_H)^2+H_5(2R(18r+23r_H)&&\nonumber\\
    &+(1+{\cal C})(13+12Rr_H)+48R^2r^2_H)]+3r^6r^3_H[(1+{\cal C})^2(10+9{\cal C}+2Rr)+4(1+{\cal C})Rr_H(19+19{\cal C}+4Rr)&&\nonumber\\
    &+4R^2r^2_H(50+50{\cal C}+8Rr)+160R^3r^3_H]+r^5r^4_H[3(1+{\cal C})^2H_4+24H_4H_5Rr_H+2H_5(1+{\cal C}+11Rr_H)]&&\nonumber\\
    &+r^4r^5_H[3H_4(1+{\cal C})^2+24H_4H_5Rr_H+2H_5(1+{\cal C}+11Rr_H)]+r^3r^6_H[8(1+{\cal C})^2+38(1+{\cal C})Rr_H+44R^2r^2_H]&&\nonumber\\
    &+6r^2r^8_HH_5R+4r^7_HH^2_5H_6,&&\nonumber\\
 u_4&=4r^8R[19(1+{\cal C})^2+98(1+{\cal C})Rr_H+ 124 R^2r^2_H]+r^7[38(1+{\cal C})^3+242 Rr_H (1+{\cal C})^2 +568 (1+{\cal C})R^2r^2_H&&\nonumber\\
    &+460R^3r^3_H]+48r^6r^2_HH_5R(1+{\cal C})+r^3r^4_H[22(1+{\cal C})^2+61Rr_H(1+{\cal C})+34R^2r^2_H]+r^2r^5_H[22(1+{\cal C})^2&&\nonumber\\
    &+61Rr_H(1+{\cal C})+2(53+36{\cal C})R^2r^2_H+240R^3r^3_H]+rr^6_H[12(1+{\cal C})^3+99(1+{\cal C})^2Rr_H+276(1+{\cal C})R^2r^2_H&&\nonumber\\
    &+240R^3r^3_H]+12r^8_HRH_5(1+{\cal C}),&&\nonumber\\
 u_5&=(r+r_H)r^9[79(1+{\cal C})^2+290(1+{\cal C})Rr_H+264R^2r^2_H]+r^9r_H[2H_5Rr_H(71+32{\cal C}+180Rr_H)]&&\nonumber\\
    &+3r^8r^2_H[2(1+{\cal C})^2(23+23{\cal C}+22Rr)+24(1+{\cal C})Rr_H(6+6{\cal C}+5Rr)+8H_5Rr_H(16+16{\cal C}+14Rr+15Rr_H)&&\nonumber\\
    &+36R^2r^2_H(3+3{\cal C}+2Rr)]+168r^7r^4_HH_4H_5R+6r^6r^5_HH_5R(13+13{\cal C}+56Rr)+3(r+r_H)r^3r^6_HH_5(9+9{\cal C} &&\nonumber\\
    &+8Rr_H)+6r^3r^8_HH_5R(1+8Rr_H)+r^2r^8_H[(1+{\cal C})^2(7+7{\cal C}+2Rr)+12Rr_H(1+{\cal C})^2+4H_5Rr_H(11+11{\cal C}&&\nonumber\\
    &+4Rr+12Rr_H)]+(r+r_H)r^9_HH_4(1+{\cal C}+4Rr_H)^2,&&\nonumber\\
 u_6&=6r^9r_HH_5R+2r^8r_H[(1+{\cal C})(7+7{\cal C}+11Rr)+Rr_H(19+19{\cal C}+26Rr)+6R^2r^2_H]+8r^7r^3_HH_4R&&\nonumber\\
    &+6r^4r^5_H[5(1+{\cal C})+9Rr_H]+6r^3r^6_H[5(1+{\cal C})+11Rr_H+6R^2r^2_H]+6r^2r^7_H[(1+{\cal C})(4+4{\cal C}+5Rr)&&\nonumber\\
    &+Rr_H(13+13{\cal C}+14Rr+6Rr_H)]+24rr^9_HH_4R+6r^{10}_HR(3+3{\cal C}+8Rr),&&\nonumber\\
 u_7&=H^5_6H_4,&&
 \end{flalign}
 \begin{flalign}
 \bar{u}_1&=r^3(H_4+2Rr_H)(1+{\cal C}+4Rr_H)^2,&&\nonumber\\
 \bar{u}_2&=2r^2[10+5R(4r+11r_H)+4R^2r_H(17r+31r_H)],&&\nonumber\\
 \bar{u}_3&=3r^9H_4[13(1+{\cal C})^2+28(1+{\cal C})Rr_H+12R^2r^2_H]+3r^8r_H[(1+{\cal C})(3{\cal C}^2+74Rr(1+{\cal C})+2Rr(19+48Rr)+4Rr_H)&&\nonumber\\
    &+32Rr_H(1+3Rr+4R^2r^2)+4Rr_H(1+{\cal C})(1+{\cal C}+30Rr)+24R^3rr^2_H]+3r^7r^2_H[16R^2r^2(5+5{\cal C}+2Rr+6Rr_H)&&\nonumber\\
    &+H_4(1+{\cal C}+4Rr)(3+3{\cal C}+4Rr+4Rr_H)]+3r^6r^3_H[8(1+{\cal C})Rr_H +20R^2r^2_H]+3r^3r^6_H(1+{\cal C})^3,&&\nonumber\\
 \bar{u}_4&=r^7[38(1+{\cal C})^2+196(1+{\cal C})Rr_H+240R^2r^2_H]+66r^6r^2_HH_5R+6r^4r^4_HR(1+{\cal C})(3H_5+1+{\cal C})+r^3r^4_H[12(1+{\cal C})^3&&\nonumber\\
    &+42Rr_H(1+{\cal C})^2+36R^2r^2_H(1+{\cal C})+96R^3r^3_H]+r^2r^5_H(1+{\cal C})[12(1+{\cal C})+15Rr_H]+rr^6_H[2(1+{\cal C})^2&&\nonumber\\
    &+14(1+{\cal C})Rr_H+20R^2r^2_H]+9r^8_HH_5 R&&,\nonumber\\
 \bar{u}_5&=2(r+r_H)r^9H_4[23(1+{\cal C})^2+72Rr_H(1+{\cal C})+54R^2r^2_H]+3r^8r^2_H[35(1+{\cal C})^2+122(1+{\cal C})Rr_H+54H_5Rr_H&&\nonumber\\
    &+104R^2r^2_H]+6r^6r^4_HH_5R(27r+13r_H)+3(r+r_H)r^3r^6_HH_4(1+{\cal C})(7+7{\cal C}+12Rr_H)+r^2r^8_H[(1+{\cal C})^2&&\nonumber\\
    &+2(1+{\cal C})Rr_H+6H_5Rr_H]+(r+r_H)r^9_HH_5 (1+{\cal C}+6Rr_H),&&\nonumber\\
 \bar{u}_6&=2r^8r_H[7(1+{\cal C})+16Rr_H]+6r^7r^3_HR+12(r+r_H)r^3r^5_HH_4[2(1+{\cal C})+3Rr_H]+6r^2r^7_H[3(1+{\cal C})+8Rr_H]&&\nonumber\\
    &+18(r+r_H)r^9_HR,&&\nonumber\\
 \bar{u}_7&=H^5_6,&&
 \end{flalign}
 \begin{flalign}
 v_1&=2r^6R(2+{\cal C}+6Rr_H)+r^5[10+66Rr_H+96R^2r^2_H+7{\cal C}^2+(1+{\cal C})(17+64Rr_H)]+r^4r_H[10+62Rr_H&&\nonumber\\
    &+120R^2r^2_H+96R^3r^3_H+{\cal C}^2(7+6Rr_H)+(1+{\cal C})(17+68Rr_H+48R^2r^2_H)]+3r^3r^2_H[32R^2r^2_H(1+{\cal C}+Rr_H)&&\nonumber\\
    &+2H^2_5+(1+{\cal C})^2(1+{\cal C}+10Rr_H)]+4r^3_HH^2_5H_6,&&\nonumber \\
 v_2&=16r^7R^2+8r^5H_4(H_5+Rr)+9r^3r^2_HH_5+r^2r^3_H[9H_5+(1+{\cal C}+4Rr_H)^2]+rr^4_H[8+2Rr_H(11+8Rr_H)&&\nonumber\\
    &+(1+{\cal C})(13+16Rr_H)+5{\cal C}^2]+3r^5_HH_5,&&\nonumber   \\
 v_3&=2r^9H_5[15+8(1+{\cal C})+54Rr_H]+2r^8r_H[2(1+{\cal C})^2(19+3Rr_H)+2(1+{\cal C})Rr_H(97+195Rr_H)+236R^2r^2_H&&\nonumber\\
    &+684R^3r^3_H]+2r^7r^2_H[45(1+{\cal C})^3+318(1+{\cal C})^2Rr_H+810(1+{\cal C})R^2r^2_H+684R^3r^3_H]+6r^6r^4_HR[20{\cal C}^2&&\nonumber\\
    &+2Rr_H(13+24Rr_H)+(1+{\cal C})(13+80Rr_H)]+6r^5r^5_HR[13{\cal C}^2+12Rr_H(1+{\cal C})]+r^3r^6_HH_5[1+4(1+{\cal C})&&\nonumber\\
    &+8Rr_H]+r^2r^7_H(5+22Rr_H+64R^2r^2_H+32R^3r^3_H)+rr^8_HH_5[(1+{\cal C})^2+8(2+{\cal C})Rr_H+16R^2r^2_H]+8r^{10}_HH_5R,&&\nonumber\\
 v_4&=6r^9H_5R(11+11{\cal C}+26Rr_H)+r^8[(1+{\cal C})^2(79+79{\cal C}+92Rr)+2(1+{\cal C})Rr_H(217+217{\cal C}+224Rr)&&\nonumber\\
    &+8R^2r^2_H(89+89{\cal C}+67Rr)+312R^3r^3_H]+80r^7r^2_HH_5R(H_4+2Rr_H)+3r^4r^4_HH_5[13+35(1+{\cal C})+60Rr_H]&&\nonumber\\
    &+3r^3r^5_H[35(1+{\cal C})^2+122(1+{\cal C})Rr_H+104R^2r^2_H+2H_5Rr_H(14+42Rr_H)]+3r^2r^6_H[22(1+{\cal C})^3&&\nonumber\\
    &+60(1+{\cal C})^2Rr_H+9(1+{\cal C})R^2r^2_H+2H_5Rr_H(43+43{\cal C}+32Rr+42Rr_H)]+96rr^8_HH_4H_5R&&\nonumber\\
    &+12r^9_HH_5R(7+7{\cal C}+16Rr),&&\nonumber\\
 v_5&=2(r+r_H)r^9r_H(7+7{\cal C}+13Rr_H)+2r^9r^3_H R(8+5{\cal C}+18Rr_H)+3r^8r^3_H[(1+{\cal C})(11+11{\cal C}+16Rr)&&\nonumber\\
    &+6Rr_H(3+3{\cal C}+4Rr)+2Rr_H(7+7{\cal C}+8Rr+6Rr_H)]+24r^7r^5_HH_4R+6r^6r^6_HR(1+{\cal C}+8Rr)&&\nonumber\\
    &+6r^4r^7_H(3+3{\cal C}+5Rr_H)+6r^3r^8_H(3+3{\cal C}+5Rr_H+2R^2r^2_H)+r^2r^9_H[(1+{\cal C})(5+5{\cal C}+4Rr)+6(1+{\cal C})Rr_H&&\nonumber\\
    &+2Rr_H(7+7{\cal C}+8Rr+6Rr_H)]+2(r+r_H)r^{10}_HH_4(1+{\cal C}+4Rr_H),&&\nonumber\\
 v_6&=rH^5_6H_4,&&
 \end{flalign}
 \begin{flalign}
 \bar{v}_1&=2r^6R{\cal C}^2+r^5[17+64Rr_H+2Rr_H{\cal C}^2+(1+{\cal C}^3)]+r^4r_H[17+68Rr_H+48R^2r^2_H+(1+{\cal C}^3)],&&\nonumber \\
 \bar{v}_2&=rr^4_H(13+16Rr_H), &&\nonumber  \\
 \bar{v}_3&=12r^{10}R[5(1+{\cal C})^2+12(1+{\cal C})Rr_H+6R^2r^2_H]+2r^9H_5(15{\cal C}^2+36(1+{\cal C})Rr_H+114R^2r^2_H)+30r^8r_H(1+{\cal C})^3&&\nonumber\\
    &+4r^7r^2_H[11(1+{\cal C})^2+71(1+{\cal C})Rr_H+98R^2r^2_H]+120r^6r^4_HR(1+4Rr_H)+6r^5r^5_HR(14+{\cal C}+54Rr_H)&&\nonumber\\
    &+6r^4r^6_H(1+{\cal C})^2+r^3r^6_HH_5[1+3(1+{\cal C})^2+6Rr_H+48R^2r^2_H]+r^2r^7_H[{\cal C}^2(5+22Rr_H)+3(1+{\cal C}^3)&&\nonumber\\
    &+(1+{\cal C})(1+4Rr_H+32R^2r^2_H)],&&\nonumber\\
 \bar{v}_4&=r^8[79(1+{\cal C})^2+368(1+{\cal C})Rr_H+420R^2r^2_H]+78r^7r^2_HH_5R+6r^5r^4_HR[(1+{\cal C})^2+12(1+{\cal C})H_5+9H^2_5]&&\nonumber\\
    &+3r^4r^4_HH_5[22(1+{\cal C})^2+60(1+{\cal C})Rr_H+13+8Rr_H+36R^2r^2_H]+3r^3r^5_H(1+{\cal C})[22(1+{\cal C})^2+46(1+{\cal C})Rr_H&&\nonumber\\
    &+12R^2r^2_H]+3r^2r^6_H[9(1+{\cal C})^2+26(1+{\cal C})Rr_H+30H_5Rr_H+16R^2r^2_H]+2r^8_HH_5R(45r+42r_H),&&\nonumber\\
 \bar{v}_5&=(r+r_H)r^9r_HH_4(11+11{\cal C}+18Rr_H)+6r^8r^3_H[5(1+{\cal C})+12Rr_H]+6r^6r^5_HR(3r+r_H)&&\nonumber\\
    &+3(r+r_H)r^3r^7_HH_4(5+5{\cal C}+6Rr_H)+2r^9_HH_6(1+{\cal C}+4Rr_H),&&\nonumber\\
 \bar{v}_6&=rH^5_6.&&
 \end{flalign}
In the limit of $R\rightarrow 0$ and ${\cal C}\rightarrow 0$,
$H_4$ and $H_5$ become unity, and thus these coefficients are
reduced to
 \begin{flalign}
 u_1&=r^3+r_HH_6,&&\nonumber\\
 u_2&=20r^2,&&\nonumber\\
 u_3&=39r^9+30r^8r_H+39r^7r^2_H+30r^6r^3_H+5r^5r^4_H+5r^4r^5_H+8r^3r^6_H+4r^7_HH_6,&&\nonumber\\
 u_4&=38r^7+22r^3r^4_H+22r^2r^5_H+12rr^6_H,&&\nonumber\\
 u_5&=79r^{10}+79r^9r_H+138r^8r^2_H+27r^4r^6_H+27r^3r^7_H+7r^2r^8_H+rr^9_H+rh^{10},&&\nonumber\\
 u_6&=14r^8r_H+30r^4r^5_H+30r^3r^6_H+24r^2r^7_H,&&\nonumber\\
 u_7&=H^5_6,&&
 \end{flalign}
 \begin{flalign}
 \bar{u}_1&=r^3,&&\nonumber\\
 \bar{u}_2&=20r^2,&&\nonumber\\
 \bar{u}_3&=39r^9+9r^7r^2_H+3r^3r^6_H,&&\nonumber\\
 \bar{u}_4&=38r^7+12r^3r^4_H+12r^2r^5_H+2rr^6_H,&&\nonumber\\
 \bar{u}_5&=46r^{10}+46r^9r_H+105r^8r^2_H+21r^4r^6_H+21r^3r^7_H+r^2r^8_H+rr^9_H+r^{10}_H,&&\nonumber\\
 \bar{u}_6&=14r^8r_H+24r^4r^5_H+24r^3r^6_H+18r^2r^7_H,&&\nonumber\\
 \bar{u}_7&=H^5_6,&&
 \end{flalign}
 \begin{flalign}
 v_1&=27r^5+27r^4r_H+9r^3r^2_H+3r^3_H H_6,&&\nonumber \\
 v_2&=8r^5+9r^3r^2_H+10r^2r^3_H+21rr^4_H+3r^5_H,&&\nonumber   \\
 v_3&=46r^9+76r^8r_H+90r^7r^2_H+5r^3r^6_H+5r^2r^7_H+rr^8_H,&&\nonumber\\
 v_4&=79r^8+144r^4r^4_H+105r^3r^5_H+66r^2r^6_H,&&\nonumber\\
 v_5&=14r^{10}r_H+14r^9r^2_H+33r^8r^3_H+18r^4r^7_H+18r^3r^8_H+5r^2r^9_H+2rr^{10}_H+2r^{11}_H,&&\nonumber\\
 v_6&=rH^5_6,&&
 \end{flalign}
 \begin{flalign}
 \bar{v}_1&=18r^5+18r^4r_H,&&\nonumber \\
 \bar{v}_2&=13rr^4_H,&& \nonumber  \\
 \bar{v}_3&=30r^8r_H+44r^7r^2_H+4r^3r^6_H+4r^2r^7_H,&&\nonumber\\
 \bar{v}_4&=79r^8+105r^4r^4_H+66r^3r^5_H+27r^2r^6_H,&&\nonumber\\
 \bar{v}_5&=11r^{10}r_H+11r^9r^2_H+30r^8r^3_H+15r^4r^7_H+15r^3r^8_H+2r^2r^9_H+2rr^{10}_H+2r^{11}_H,&&\nonumber\\
 \bar{v}_6&=rH^5_6.&&
 \end{flalign}
Thus, after replacing $H_6$ with $H_0$ in the massless limit, the
differences of $u_i-\bar{u}_i~(i=1,2,\cdot\cdot\cdot,7)$ and
$v_i-\bar{v}_i~(i=1,2,\cdot\cdot\cdot,6)$ can be obtained as
 \begin{flalign}\label{diffu}
 u_1-\bar{u}_1&=r_HH_0,&&\nonumber\\
 u_2-\bar{u}_2&=0,&&\nonumber\\
 u_3-\bar{u}_3&=r_HH_0(30r^6+5r^3r^3_H+4r^6_H),&&\nonumber\\
 u_4-\bar{u}_4&=10rr^4_HH_0,&&\nonumber\\
 u_5-\bar{u}_5&=r^2H_0(33r^6+6r^6_H),&&\nonumber\\
 u_6-\bar{u}_6&=6r^2r^5_HH_0,&&\nonumber\\
 u_7-\bar{u}_7&=0,&&
 \end{flalign}
 \begin{flalign}
 \label{diffv}
 v_1-\bar{v}_1&=H_0(9r^3+4r^3_H),&&\nonumber\\
 v_2-\bar{v}_2&=8r^5+9r^3r^2_H+10r^2r^3_H+8rr^4_H+3r^5_H,&&\nonumber\\
 v_3-\bar{v}_3&=rH_0(46r^6+r^6_H),&&\nonumber\\
 v_4-\bar{v}_4&=39r^2r^4_HH_0,&&\nonumber\\
 v_5-\bar{v}_5&=3r^2r_HH_0(r^6+r^6_H),&&\nonumber\\
 v_6-\bar{v}_6&=0.&&
 \end{flalign}
which are the exactly same coefficients in $z^5$ and $z^6$ of the
HRBH in massless gravity in Eqs. (\ref{gems-Haywardmassless}),
respectively.

\end{document}